\documentclass[aps,showpacs,floats,floatfix]{revtex4}
\usepackage{graphicx,axodraw_3}

\thicklines
\begin{document}
\def\slash#1{#1 \hskip -0.5em / } 
\def\beq{\begin{equation}}
\def\eeq{\end{equation}}
\def\beqy{\begin{eqnarray}}
\def\eeqy{\end{eqnarray}}

\thispagestyle{empty}
\preprint{JLAB-THY-04-38}

\title{Polarization Observables in $\gamma N\to K\overline{K}N$}
\author{
W. Roberts}
\affiliation{Department of Physics, Old Dominion University, Norfolk, VA 23529,
USA \\
and \\
Continuous Electron Beam Accelerator Facility \\
12000 Jefferson Avenue, Newport News, VA 23606, USA.}

\date{\today} 

\begin{abstract} 
We explore some of the rich structure of the polarization observables recently
developed for processes like $\gamma N\to\pi\pi N$ and $\gamma N\to K
\overline{K} N$ in the framework of a specific model for the latter process.
Emphasis is placed on observables that may be accesible at existing facilities
in the near future. The sensitivity of the observables to the details of the
model indicate that they will be a very useful tool in differentiating between
different models for reactions like these. In the framework of a model for
$\gamma N\to K \overline{K} N$, we examine the sensitivity of the observables
to coupling constants of the $\phi$, to the properties of the $\Lambda(1405)$,
and to the existence of the $\Theta^+$.
\flushright{JLAB-THY-04-XXX} 
\end{abstract}
\maketitle 
\setcounter{page}{1}

\section{Introduction and Motivation}

In a recent article \cite{polarization}, we introduced sets of polarization observables
for the processes $\pi N\to\pi\pi N$ and $\gamma N\to\pi\pi N$. In this note, we
examine some of these observables in the context of a specific model for the process 
$\gamma N\to K \overline{K}N$. We explore four different facets of these polarization
observables. First, we examine the expected sizes of these observables. Obviously, if
these observables are too small, they may not be of interest to experimentalists. It
turns out that some of the observables, including some that can be accessed in the near
future at present facilities, can be quite large.

We also explore the sensitivity of the observables to the details of the
underlying dynamics, thus illustrating how they will be useful in helping to
pin down parameters in any model used to describe such processes. In addition,
since these observables are five-fold differential observables, we examine
various ways of presenting them, noting that the same observable can appear
very different depending on what is chosen as the independent kinematic
variables. Finally, we explore the potential of these observables in the hunt
for resonances, by examining them in a model with and without the $\Theta^+$.

We focus on the 16 observables that may be readily measured at present
facilities like Jefferson Lab, Bonn and Graal, for example. The availability of
linearly or circularly polarized beams at these facilities, along with advances
in the technology for the production of polarized targets, means that a number
of these observables can, in principle, be measured with high precision.
Indeed, first measurements of $I^\odot$, the beam asymmetry that arises with
circularly polarized photons, in $\gamma p\to p\pi^+\pi^-$ indicate that even
the smaller asymmetries can be measured with good precision\cite{strauch}.
Triple polarization observables can be measured from the self-analysing decays
of hyperons produced in  processes like $\gamma N\to\pi K Y$. However, such
observables will be more difficult to measure for processes like $\gamma
N\to\pi\pi N$, as none of the present facilities are equiped to measure recoil
polarizations.

The rest of this note is organized as follows. In the next section, we present
the kinematics for the process, as well as the observables. We also briefly
discuss the model that we use to calculate the observables in section II. In
section III, we present our results, by examining the sensitivity of some of
the observables to (a) the coupling constants of the $\phi$ meson; (b) the
$\Lambda(1405)$; and (c) the $\Theta^+$, including its parity. 
Section IV presents our conclusions and an
outlook.

\section{Kinematics, Observables and Model}

\subsection{Kinematics}

For the process $\gamma N\to K\overline{K}N$, momentum conservation gives
\beq
p_1+k=p_2+q_1+q_2,
\eeq
where $k$ is the momentum of the incident photon, $p_1$ is that of the target
nucleon, $p_2$ is that of the recoil nucleon, $q_1$ is the momentum of the kaon,
and $q_2$ is the momentum of the $\overline{K}$. The momentum of the incident
photon is chosen to define the $z$-axis, with
\begin{equation}\label{phok}
k=\left(\omega,0,0,K\right)\equiv \left(\omega,{\bf K}\right),
\end{equation}
and
\begin{equation}\label{nucp}
p_1=\left(\sqrt{s}-\omega,0,0,-K\right),
\end{equation}
where
\begin{equation} 
s=(k+p_1)^2
\end{equation}
is the square of the total center-of-mass (cm) energy,
\begin{equation}
K=\omega=\frac{s-m^2}{2\sqrt{s}},
\end{equation}
and $m$ is the mass of the nucleon.

For the three final-state particles, we define
\begin{equation}\label{kinepnppn1}
p_2=\left(\frac{s+m_N^2-s_{K\overline{K}}}{2\sqrt{s}},
-Q\sin{\theta},0,-Q\cos{\theta}\right),
\end{equation}
where 
\begin{equation}
s_{K\overline{K}}=\left(q_1+q_2\right)^2
\end{equation}
and
\begin{equation}
Q=\frac{\lambda^{1/2}\left(s,s_{K\overline{K}},m^2\right)}{2\sqrt{s}}.
\end{equation}
Here, we are using the recoiling nucleon, or more precisely, the recoiling 
pair of kaons, to define the collision plane. 
The momentum of the kaon may be written
\beq\label{kinepnppn2}
\vec{q}_1=Q_1\left(\sin{\theta_1}\cos{\phi_1},\sin{\theta_1}\sin{\phi_1},\cos{\theta_1}\right),
\eeq
where $Q_1$, $\theta_1$ and $\phi_1$ can be written in terms of $s$,
$s_{K\overline{K}}$, $\theta$, and the angles describing the motion of the pair of
kaons in their cm frame. For the purposes of this discussion, it is more
appropriate to discuss the momenta of the kaons in their rest frame. In this
case, we choose a $z^\prime$ axis to be along the momentum of the recoiling
nucleon. Relative to this axis, in their cm frame, the momenta of the kaons are
\beqy
q_1^*&=&\left(\sqrt{s_{K\overline{K}}}/2,Q^*\sin{\Theta^*}\cos{\Phi^*},
Q^*\sin{\Theta^*}\sin{\Phi^*},Q^*\cos{\Theta^*}\right),\nonumber\\
q_2^*&=&\left(\sqrt{s_{K\overline{K}}}/2,-Q^*\sin{\Theta^*}\cos{\Phi^*},
-Q^*\sin{\Theta^*}\sin{\Phi^*},-Q^*\cos{\Theta^*}\right),
\eeqy
where
\beq
Q^*=\frac{\sqrt{s_{K\overline{K}}-4m_K^2}}{2},
\eeq
and the asterisk on a quantity indicates that it refers to the rest frame of the
pair of kaons.

\subsection{Observables}

We can write the reaction rate $I$, for the process $\gamma N\to
K\overline{K}N$, as
\beqy
\rho_fI&=&I_0\left\{\left(1+\vec{\Lambda}_i\cdot\vec{P}+\vec{\sigma}\cdot\vec{P}^\prime+
\Lambda_i^\alpha\sigma^{\beta^\prime}{\cal O}_{\alpha\beta^\prime}\right)\right.\nonumber\\
&&+\left.\delta_\odot\left(I^\odot+\vec{\Lambda}_i\cdot\vec{P}^\odot+\vec{\sigma}\cdot\vec{P}^{\odot\prime}+
\Lambda_i^\alpha\sigma^{\beta^\prime}{\cal O}^\odot_{\alpha\beta^\prime}\right)\right.\nonumber\\
&&+\left.\delta_\ell\left[\sin{2\beta}\left(I^s+\vec{\Lambda}_i\cdot\vec{P}^s+\vec{\sigma}\cdot\vec{P}^{s\prime}+
\Lambda_i^\alpha\sigma^{\beta^\prime}{\cal O}^s_{\alpha\beta^\prime}\right)\right.\right.\nonumber\\
&&+\left.\left.\cos{2\beta}\left(I^c+\vec{\Lambda}_i\cdot\vec{P}^c+\vec{\sigma}\cdot\vec{P}^{c\prime}+
\Lambda_i^\alpha\sigma^{\beta^\prime}{\cal
O}^c_{\alpha\beta^\prime}\right)\right]\right\},
\eeqy
where $\vec{P}$ represents the polarization asymmetry that arises if the target
nucleon has polarization $\vec{\Lambda}_i$, $\rho_f=\frac{1}{2}\left(1+\vec{\sigma}\cdot
\vec{P}^\prime\right)$ is the density matrix of the recoiling nucleon, and
${\cal O}_{\alpha\beta^\prime}$ is the observable if both the target and recoil
polarization are measured. The primes indicate that the recoil observables are
measured with respect to a set of axes $x^\prime,\,y^\prime,\,z^\prime$, in
which $z^\prime$ is along the direction of motion of the recoiling nucleon, and
$y^\prime=y$. $\delta_\odot$ is the degree of circular polarization in the
photon beam, while  $\delta_\ell$ is the degree of linear polarization, with
the direction of polarization being at an angle $\beta$ to the $x$-axis.

Of these 63 polarization observables ($I_0$ is proportional to the unpolarized
differential cross section), 48 require detection of the polarization of the
recoil nucleon. Since none of the present facilities are equiped for such
measurements, we devote little attention to such observables at this time. We
note, however, such observables will be essential for extracting the helicity or
transversity amplitudes that describe this process.

\subsection{Model}

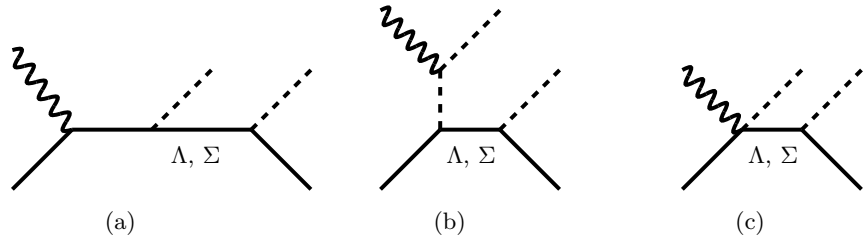
\begin{figure}
\begin{center}
\begin{picture}(100,50)
\Line(0,0)(15,15)1
\Line(15,15)(60,15)1
\Line(60,15)(75,0)1
\Photon(0,35)(15,15)251
\DashLine(60,15)(75,30)21
\DashLine(35,15)(50,30)21
\put(60,10){$\Lambda$, $\Sigma$}
\put(35,-15){(a)}
\end{picture}
\hskip .5in
 \begin{picture}(75,75)
\Line(0,0)(15,15)1
\Line(15,15)(30,15)1
\Line(30,15)(45,0)1
\Photon(0,45)(15,30)251
\DashLine(15,15)(15,30)21
\DashLine(15,30)(30,45)21
\DashLine(30,15)(45,30)21
\put(25,10){$\Lambda$, $\Sigma$}
\put(20,-15){(b)}
\end{picture}
\hskip .5in
\begin{picture}(75,50)
\Line(0,0)(15,15)1
\Line(15,15)(30,15)1
\Line(30,15)(45,0)1
\Photon(0,30)(15,15)251
\DashLine(15,15)(30,30)21
\DashLine(30,15)(45,30)21
\put(25,10){$\Lambda$, $\Sigma$}
\put(20,-15){(c)}
\end{picture}
\end{center}

\caption{`Born' diagrams: continuous, unlabeled lines are nucleons. Unlabeled dashed lines
are kaons and wavy lines are photons.\label{fig2}}
\end{figure}

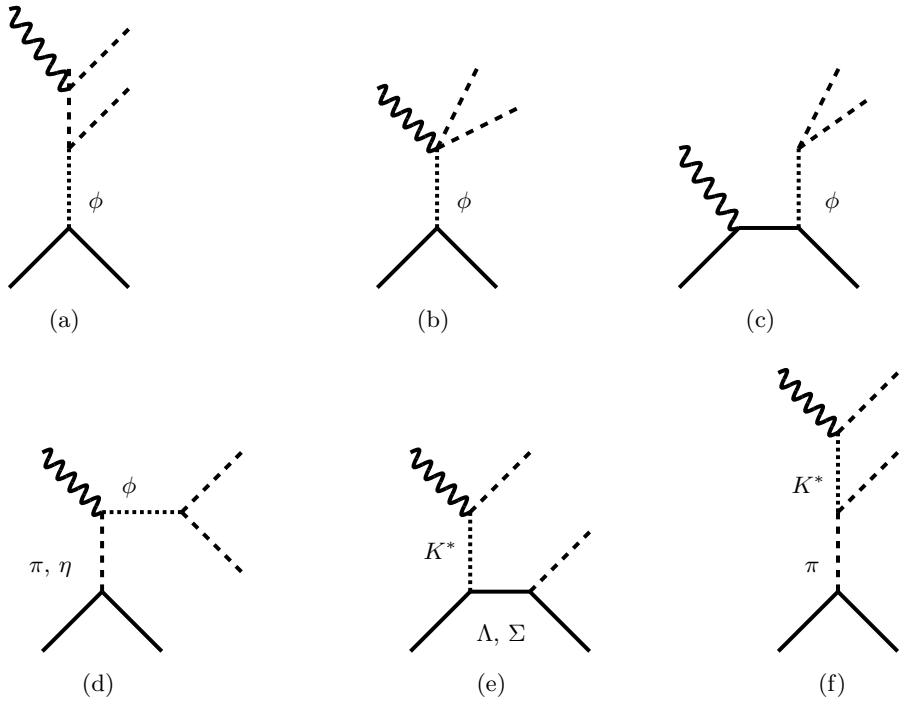
\begin{figure}
\begin{center}

\begin{picture}(100,150)
\Line(0,0)(15,15)1
\Line(15,15)(30,0)1
\DashLine(15,15)(15,35)11
\Photon(0,70)(15,50)251
\DashLine(15,35)(15,55)21
\DashLine(15,35)(30,50)21
\DashLine(15,50)(30,65)21
\put(15,-15){(a)}
\put(30,30){$\phi$}
\end{picture}
\hskip .5in
 \begin{picture}(75,75)
\Line(0,0)(15,15)1
\Line(15,15)(30,0)1
\DashLine(15,15)(15,35)11
\DashLine(15,35)(35,45)21
\DashLine(15,35)(25,55)21
\Photon(0,50)(15,35)251
\put(15,-15){(b)}
\put(30,30){$\phi$}
\end{picture}
\hskip 0.5in
\begin{picture}(100,50)
\Line(0,0)(15,15)1
\Photon(0,35)(15,15)251
\Line(15,15)(30,15)1
\Line(30,15)(45,0)1
\DashLine(30,15)(30,35)11
\DashLine(30,35)(40,55)21
\DashLine(30,35)(47,47)21
\put(55,30){$\phi$}
\put(25,-15){(c)}
\end{picture}
\vskip 1.2in
\begin{picture}(100,50)
\Line(0,0)(15,15)1
\DashLine(15,15)(15,35)21
\Line(15,15)(30,0)1
\Photon(0,50)(15,35)251
\DashLine(15,35)(35,35)11
\DashLine(35,35)(50,50)21
\DashLine(35,35)(50,20)21
\put(30,60){$\phi$}
\put(-5,30){$\pi$, $\eta$}
\put(15,-15){(d)}
\end{picture}
\hskip 0.5in
\begin{picture}(100,50)
\Line(0,0)(15,15)1
\DashLine(15,15)(15,35)11
\Photon(0,50)(15,35)251
\DashLine(15,35)(30,50)21
\Line(15,15)(30,15)1
\Line(30,15)(45,0)1
\DashLine(30,15)(45,30)21
\put(25,-15){(e)}
\put(5,35){$K^*$}
\put(25,4){$\Lambda$, $\Sigma$}
\end{picture}
\hskip 0.5in
\begin{picture}(50,50)
\Line(0,0)(15,15)1
\Line(15,15)(30,0)1
\DashLine(15,15)(15,35)21
\DashLine(15,35)(15,55)11
\DashLine(15,35)(30,50)21
\Photon(0,70)(15,55)251
\DashLine(15,55)(30,70)21
\put(15,-15){(f)}
\put(5,60){$K^*$}
\put(10,30){$\pi$}

\end{picture}
\vskip .25in
\caption{ Diagrams containing vector mesons. The dotted lines represent 
the vector mesons.\label{fig3}}
\end{center}
\end{figure}

\begin{figure}
\begin{center}

\hskip .4in
\begin{picture}(100,80)
\Line(0,0)(15,15)1
\Line(15,15)({37.5},15)1
\Line({37.5},15)(60,15){2.5}
\Line(60,15)(75,0)1
\Photon(0,35)(15,15)251
\DashLine(60,15)(75,30)21
\DashLine(35,15)(50,30)21
\put(35,-5){(a)}
\end{picture}
\hskip 1.25in
 \begin{picture}(50,50)
\Line(0,0)(15,15)1
\Line(15,15)(30,15){2.5}
\Line(30,15)(45,0)1
\Photon(0,45)(15,30)251
\DashLine(15,15)(15,30)21
\DashLine(15,30)(30,45)21
\DashLine(30,15)(45,30)21
\put(20,-5){(b)}
\end{picture}
\vskip 1.2in
 \begin{picture}(75,0)
\Line(0,0)(15,15)1
\Line(15,15)(30,15){2.5}
\Line(30,15)(45,0)1
\Photon(0,30)(15,15)251
\DashLine(15,15)(30,30)21
\DashLine(30,15)(45,30)21
\put(25,-10){(c)}
\end{picture}
\hskip 1.25in
\begin{picture}(50,0)
\Line(0,0)(15,15)1
\Line(15,15)({37.5},15){2.5}
\Line({37.5},15)(60,15){2.5}
\Line(60,15)(75,0)1
\Photon(25,35)(40,15)251
\DashLine(60,15)(75,30)21
\DashLine(15,15)(30,30)21
\put(35,-5){(d)}
\end{picture}
\end{center} 

\caption{Diagrams containing excited baryons. In (a) to (d), the
thick solid lines may be either $\Lambda^*$, $\Sigma^*$ or $\Theta$, while the
thin solid line is a nucleon. In diagram (d), the photon couples to the charge
of the intermediate resonance: in this model, we neglect couplings to any higher
electromagnetic moments of the resonance.\label{fig4}} 
\end{figure}
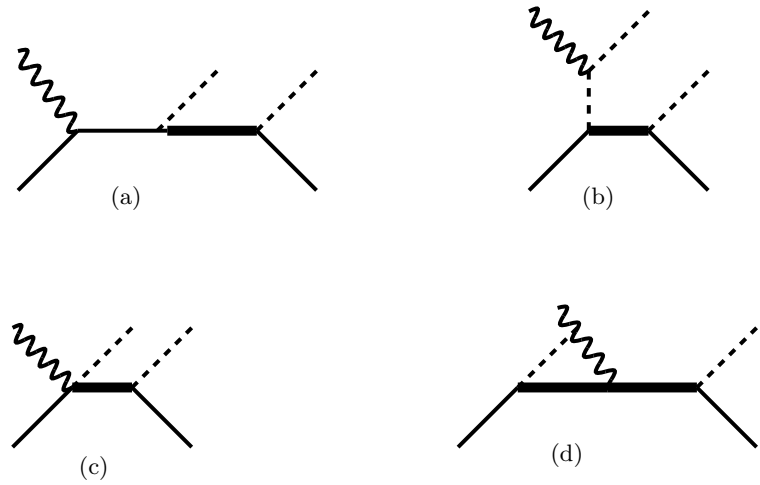

The model we use to examine these observables is one that has recently been used
to examine possible production mechanisms for the $\Theta^+$ \cite{zplus}. The model is
constructed in the framework of a phenomenological Lagrangian, and details are
given in \cite{zplus}. The ingredients of the model are summarized by the
diagrams shown in figs. \ref{fig2} to \ref{fig4}. Briefly, a number of resonant and non-resonant
contributions are included. The non-resonant contributions arise from a number
of Born terms (shown in fig \ref{fig2}), as well as diagrams with resonances
that occur in crossed channels.

The resonant contributions (shown in figs \ref{fig3} and \ref{fig4}) include
the $\phi(1020)$, along with a number of excited hyperons that appear in the
$N\overline{K}$ system. The dominant
contribution from among the hyperons arises from the $\Lambda(1520)$. We also
include the $\Theta^+$ as a resonance in the $NK$ channel. For the purposes of
this study, we take its spin to be 1/2, its mass to be 1540 MeV, and its width
to be 10 MeV. However, we allow its parity to be either positive or negative,
and examine the effects on the observables. For the results that we show, we
assume that $w=2.5$ GeV.

\section{Results}

The polarization observables for this process are five-fold differential, which
means that there are a number of different ways in which they can be displayed.
Since it is not obvious how to display five-fold differential quantities, it is
usual to integrate over some of the independent variables. In the following
subsections, we integrate over some of the kinematic variables, showing the
resulting observables as two-dimensional surfaces. 

We note that since the observables are either even or odd under the
transformation $\Phi^*\leftrightarrow 2\pi-\Phi^*$, we do not integrate over
this variable. Thus, in all the plots that follow, the axis on the `lower left'
is $\Phi^*$ (ranging from 0 to $2\pi$). Observables that are even in the
$\Phi^*$ transformation could also be presented as Dalitz plots.

The polarization observables all have values that lie between 1 and -1, and the
colors of the surfaces reflect the values of the observables. In the surfaces,
red corresponds to large positive values (near +1), while blue corresponds to
large negative values. Values near zero are green. In each of the plots, the
point of view looks slightly downward onto the surface, from a corner that
looks along the axes of the two independent variables. Along the $\Phi^*$ axis
(always on the left of the figures), the line of view is from larger values of
$\Phi^*$ (closest to the observer) to smaller values of $\Phi^*$ (furthest from
the observer). For the second independent axis, smaller values of the variable 
are closer to the observer.

\subsection{Presentation}

\begin{figure}
\caption{The observable $P_x^\odot$ shown in terms of different kinematic variables. (a): as a function
of $m_{K\overline{K}}$ and $\Phi^*$; (b): as a function of $m_{NK}$ and $\Phi^*$; (c): 
as a function of $m_{N\overline{K}}$ and $\Phi^*$; (d): as a function of $\cos{\theta}$ and 
$\Phi^*$.\\ In each of the plots, the
point of view looks slightly downward onto the surface, from a corner that
looks along the axes of the two independent variables. Along the $\Phi^*$ axis
(always on the left of the figures), the line of view is from larger values of
$\Phi^*$ (closest to the observer) to smaller values of $\Phi^*$ (furthest from
the observer). For the second independent axis, smaller values of the variable 
are closer to the observer.\\ In the surfaces,
red corresponds to large positive values (near +1), while blue corresponds to
large negative values. Values near zero are green.\label{presentation}}
\vskip 0.2in
\centerline{(a)\includegraphics[height=3.0in,angle=0]{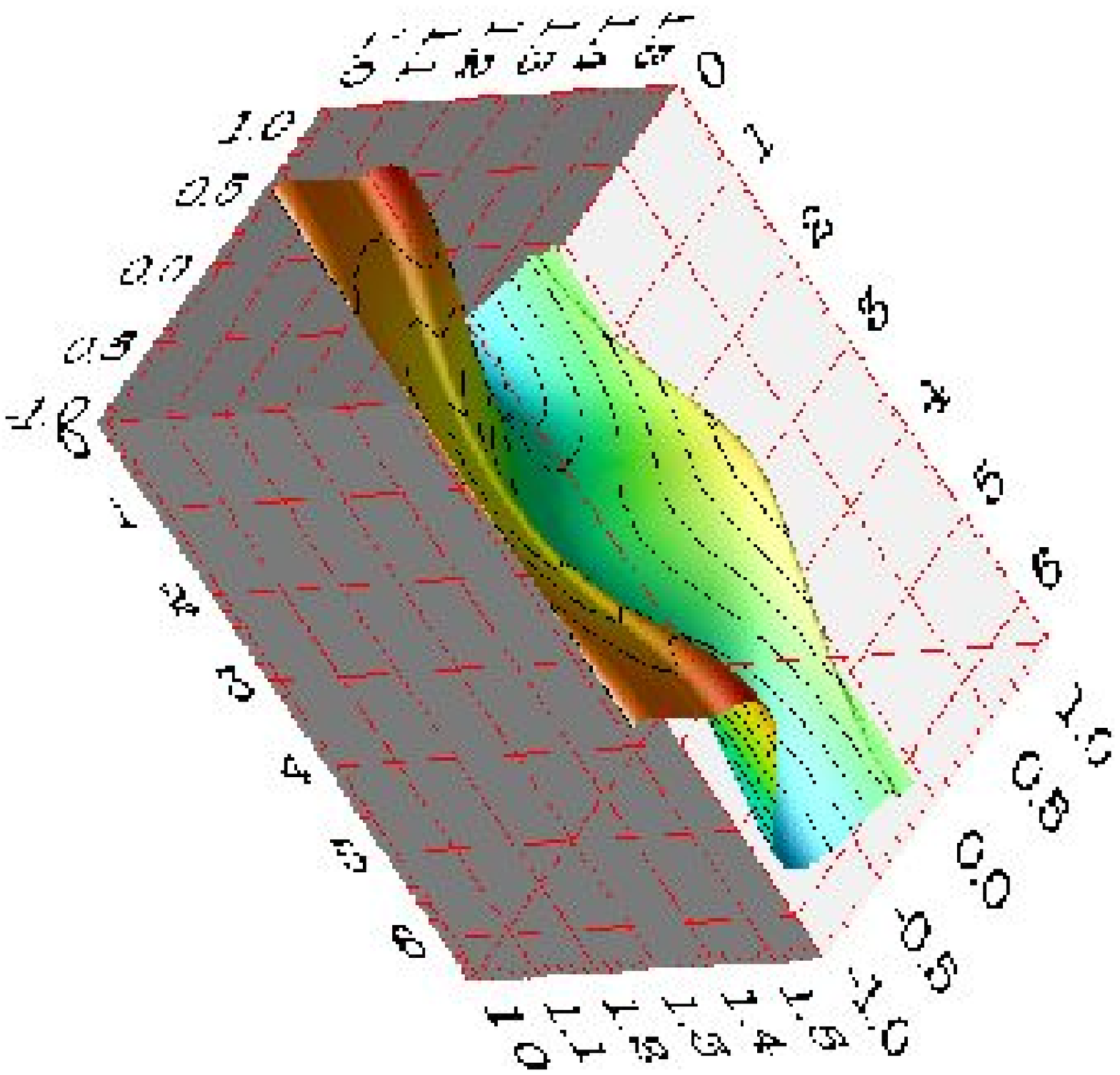}
(b)\includegraphics[height=3.0in,angle=0]{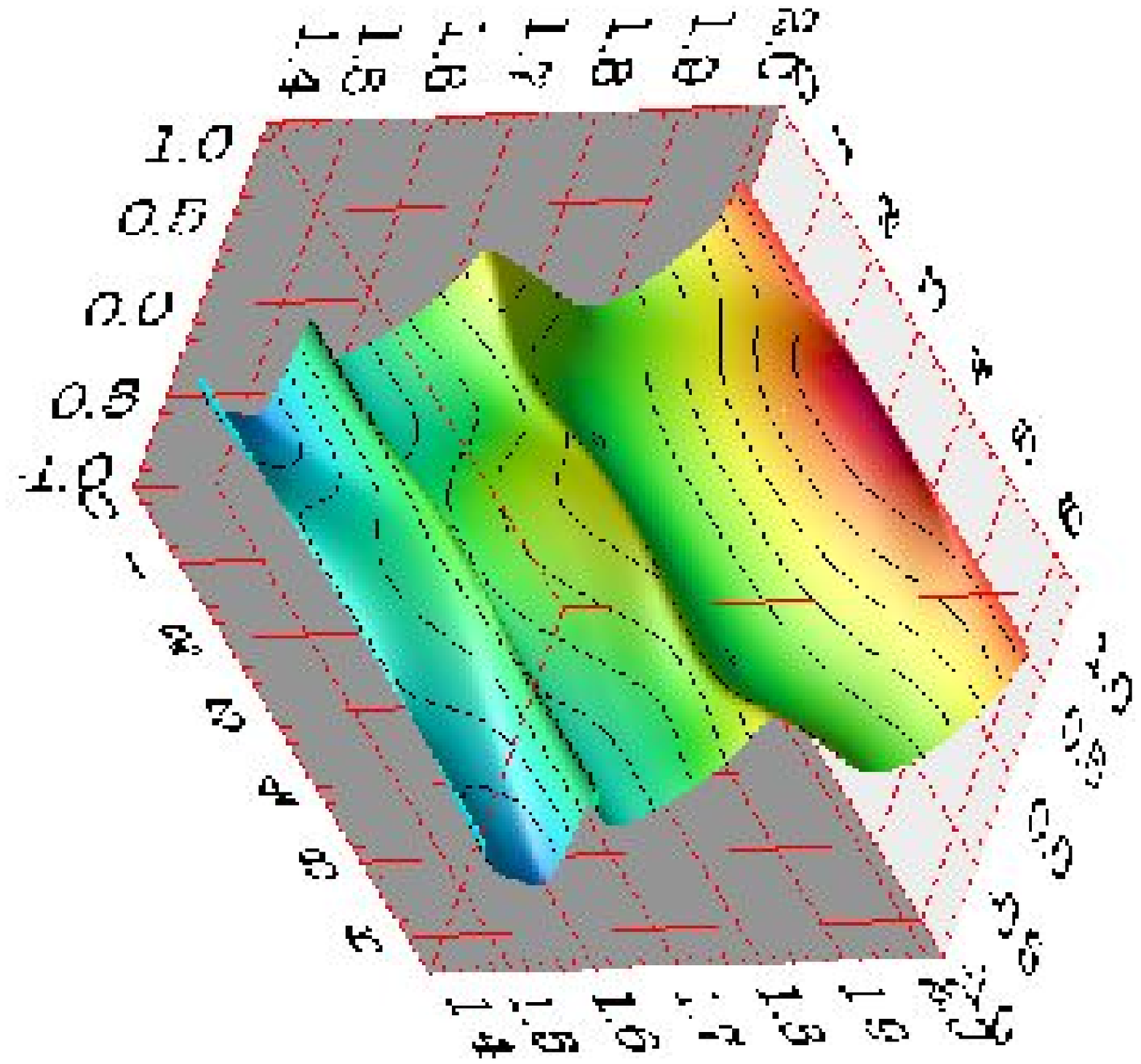}}
\vskip0.2in
\centerline{(c)\includegraphics[height=3.0in,angle=0]{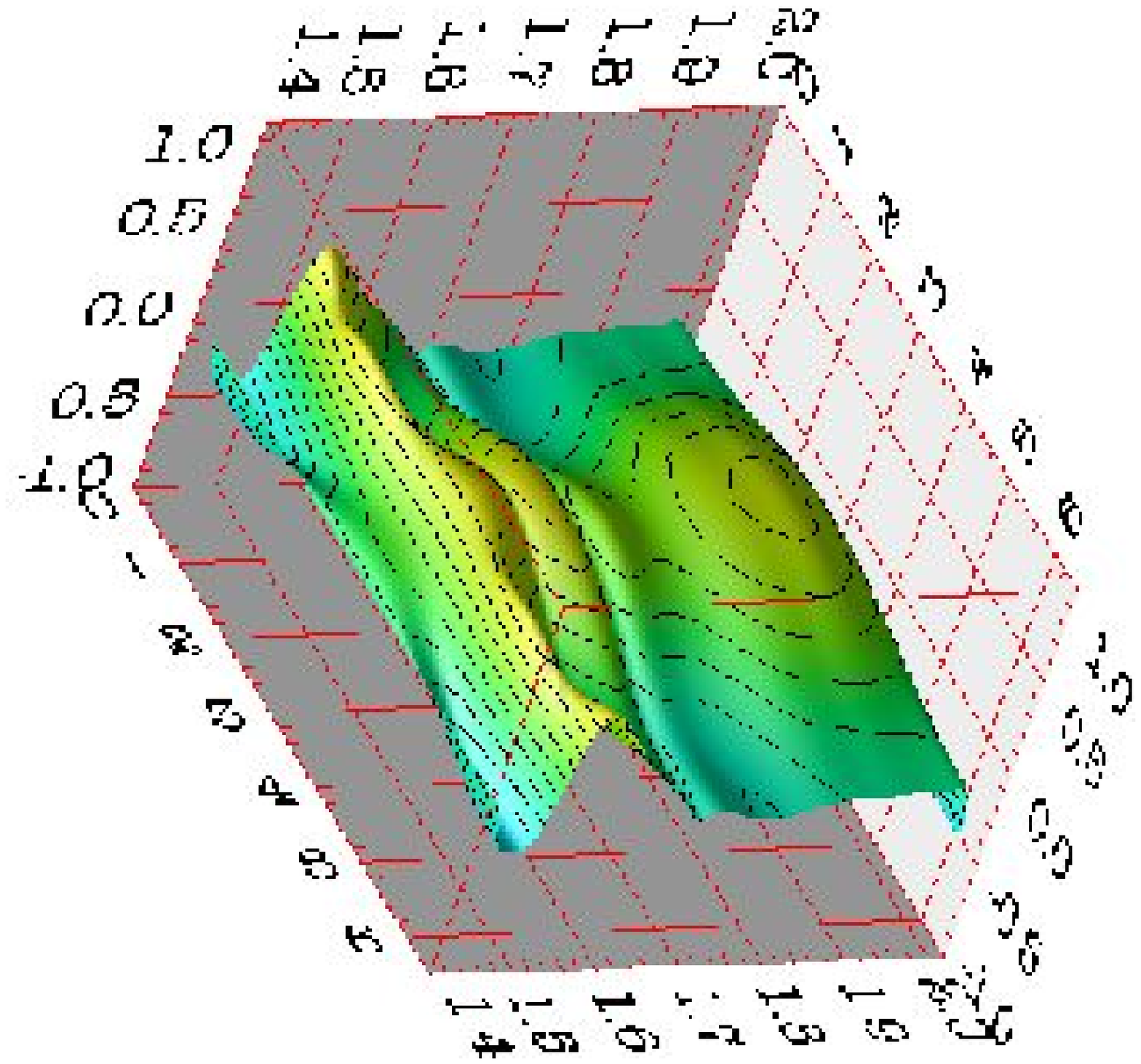}
(d)\includegraphics[height=3.0in,angle=0]{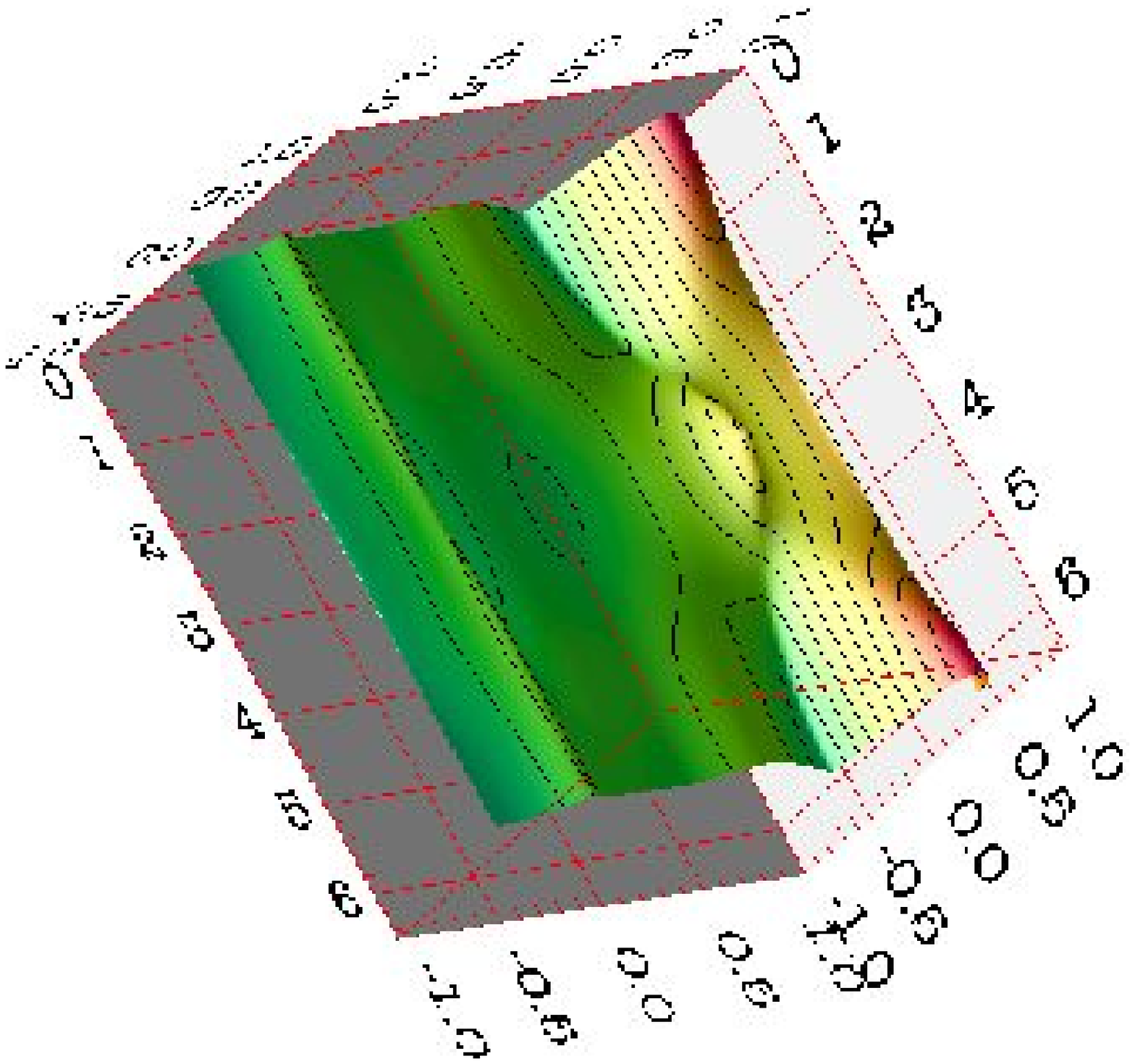}}
\end{figure}

Figure \ref{presentation} shows the observable $P_x^\odot$ (the asymmetry that arises from circularly
polarized photons incident on nucleons polarized along the $x$-axis), presented in four different ways.
In all cases, one of the independent variables is $\Phi^*$. For the surface in (a) (top left) of this
figure, the second independent variable is $m_{K\overline{K}}$ (along the horizontal axis). In (b), the
second independent variable is $m_{NK}$. For (c), it is $m_{N\overline{K}}$, while for (d), it is
$\cos{\theta}$. In these four surfaces, the content of the model is exactly the same (the process is
$\gamma p\to nK^+ \overline{K}^0$), but the appearance of the observable is very different in each case.
Note the effects of the $\Lambda(1520)$ near its resonant mass of 1.52 GeV in (c). Similarly, the effect
of the $\Theta^+$ is seen near its mass of 1.54 GeV in (b).

In (b) and (c), it might be more useful to use a different definition of
$\Phi^*$, one appropriate to the pair of hadrons whose invariant mass is
treated as the other independent variable. Thus, in (b), instead of $\Phi^*$
for the  $K\overline{K}$ system (as defined earlier), it might be more
appropriate to use $\Phi^*_{NK}$, while for (c), $\Phi^*_{N\overline{K}}$ might
be more appropriate. The reason is that if there is a resonance in the
$K\overline{K}$ system, its decay products will yield a
$\Phi^*_{N\overline{K}}$ distribution that characterizes the decay. In the same
manner, a resonance in the $NK$ ($N\overline{K}$) system should yield a
$\Phi^*_{NK}$ ($\Phi^*_{N\overline{K}}$) distribution characteristic of its 
decay to $NK$ ($N\overline{K}$).

\subsection{$\Phi$ Coupling Constants}

\begin{figure}
\caption{The observable $I^s$, showing its sensitivity to the coupling constants $G_v^\phi$ and 
$G_t^\phi$. (a): $G_v^\phi=4,\,\, G_t^\phi=0$; (b): $G_v^\phi=-4,\,\, 
G_t^\phi=0$; (c): $G_v^\phi=0,\,\, G_t^\phi=4$; (d): $G_v^\phi=0,\,\, 
G_t^\phi=-4$. In each case, the surface is shown as a function of $m_{K\overline{K}}$ and $\Phi^*$.\label{phia}}
\vskip 0.2in
\centerline{(a)\includegraphics[height=3.0in,angle=0]{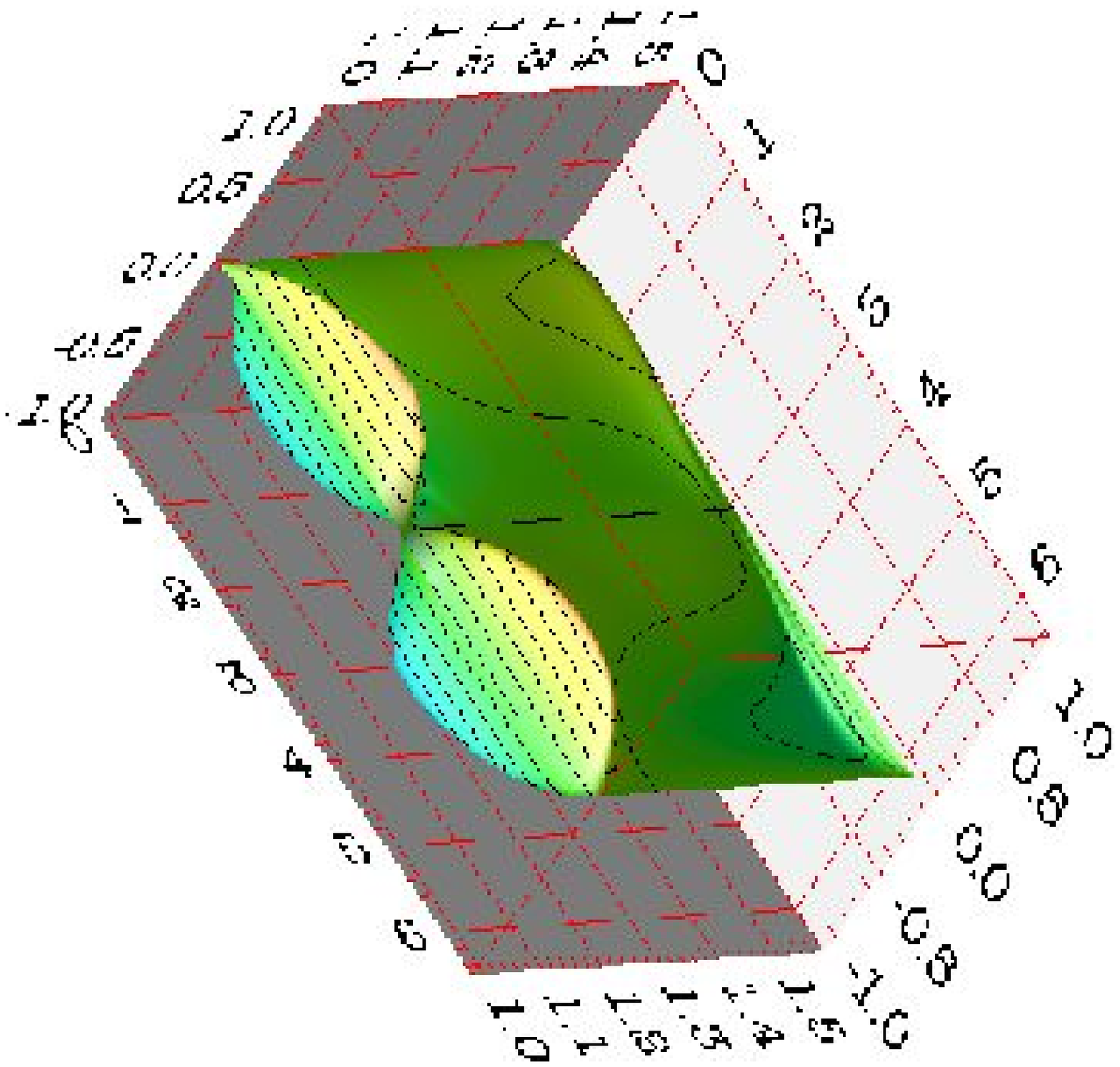}
(b)\includegraphics[height=3.0in,angle=0]{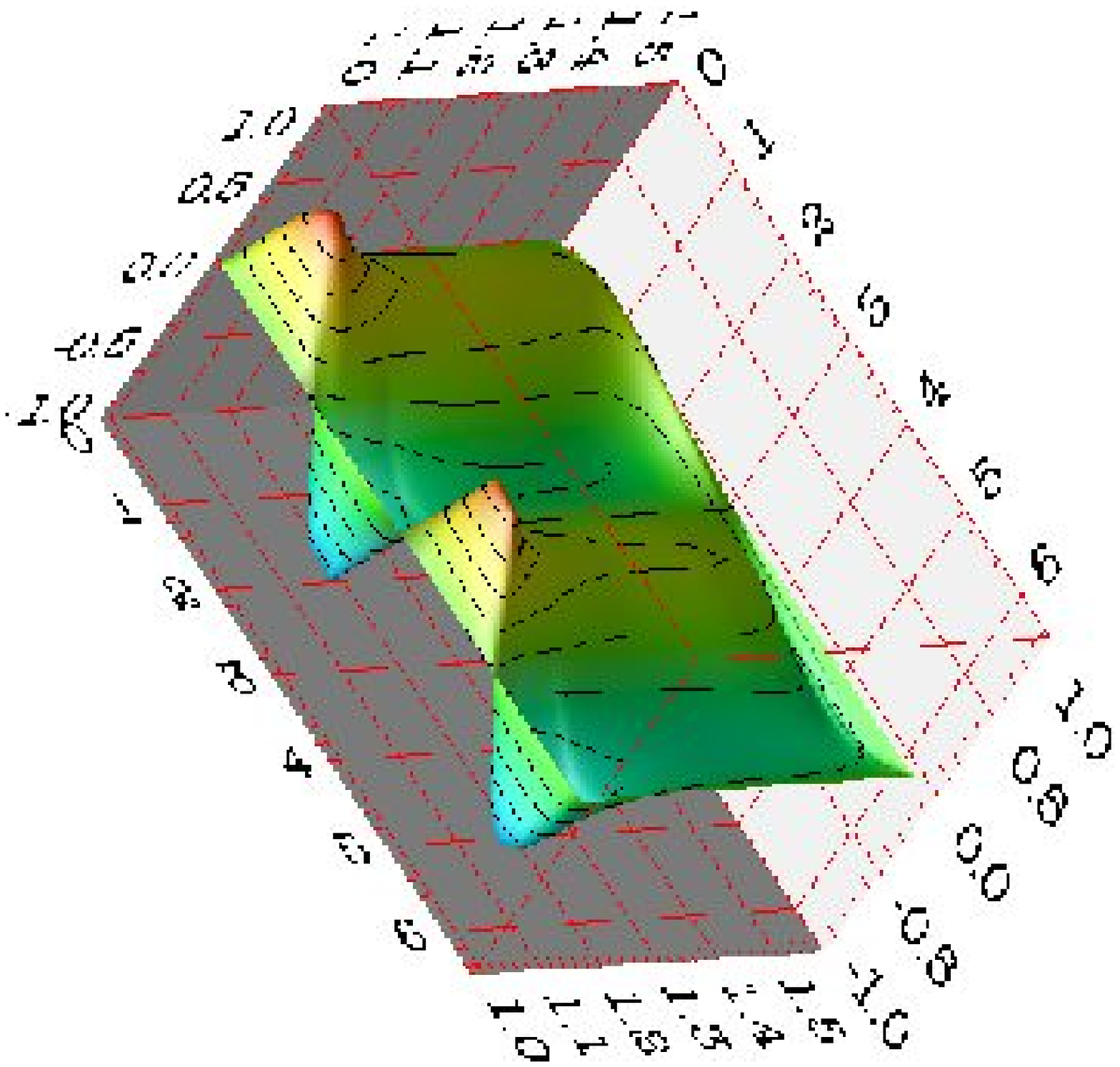}}
\vskip0.2in
\centerline{(c)\includegraphics[height=3.0in,angle=0]{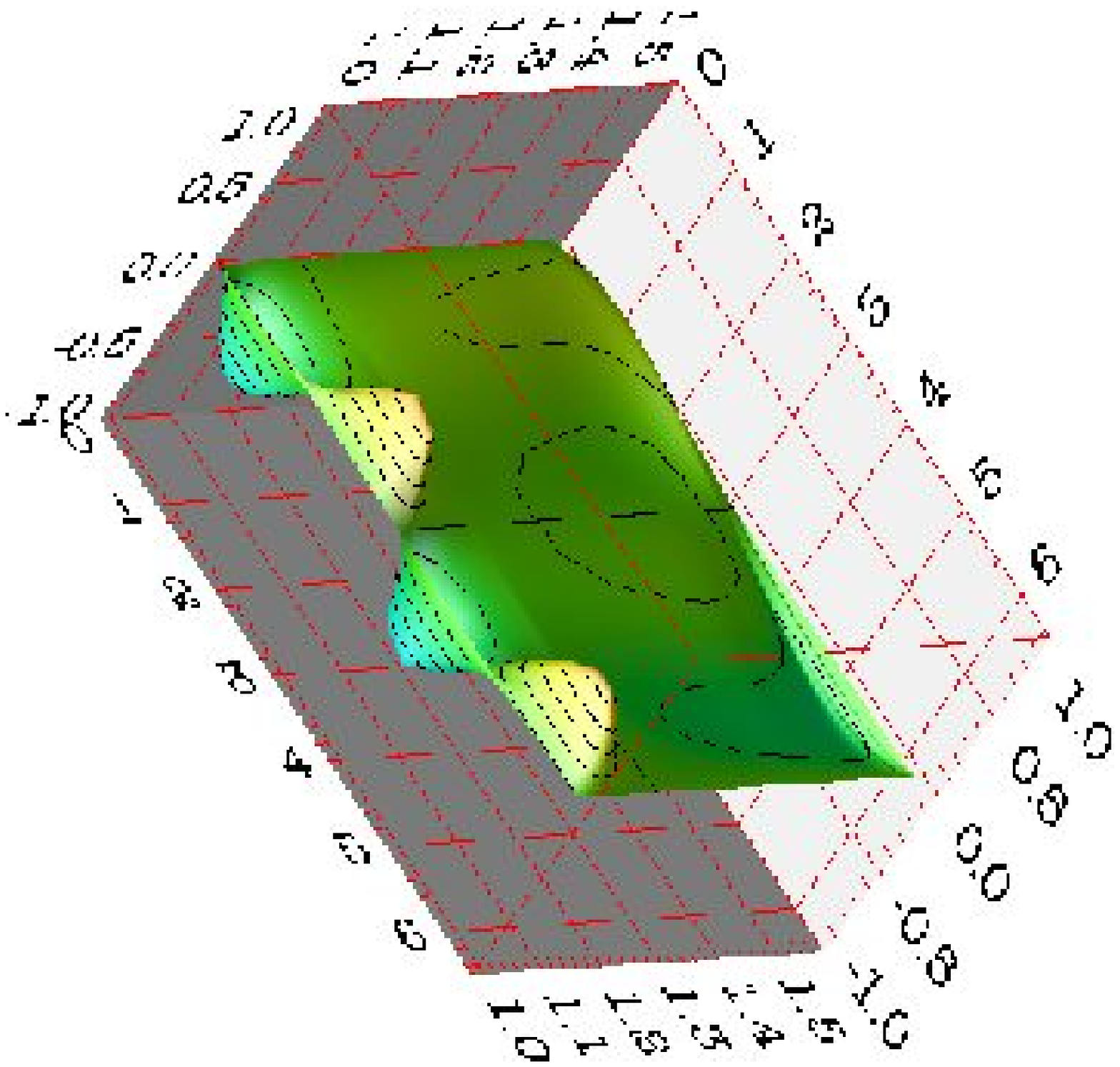}
(d)\includegraphics[height=3.0in,angle=0]{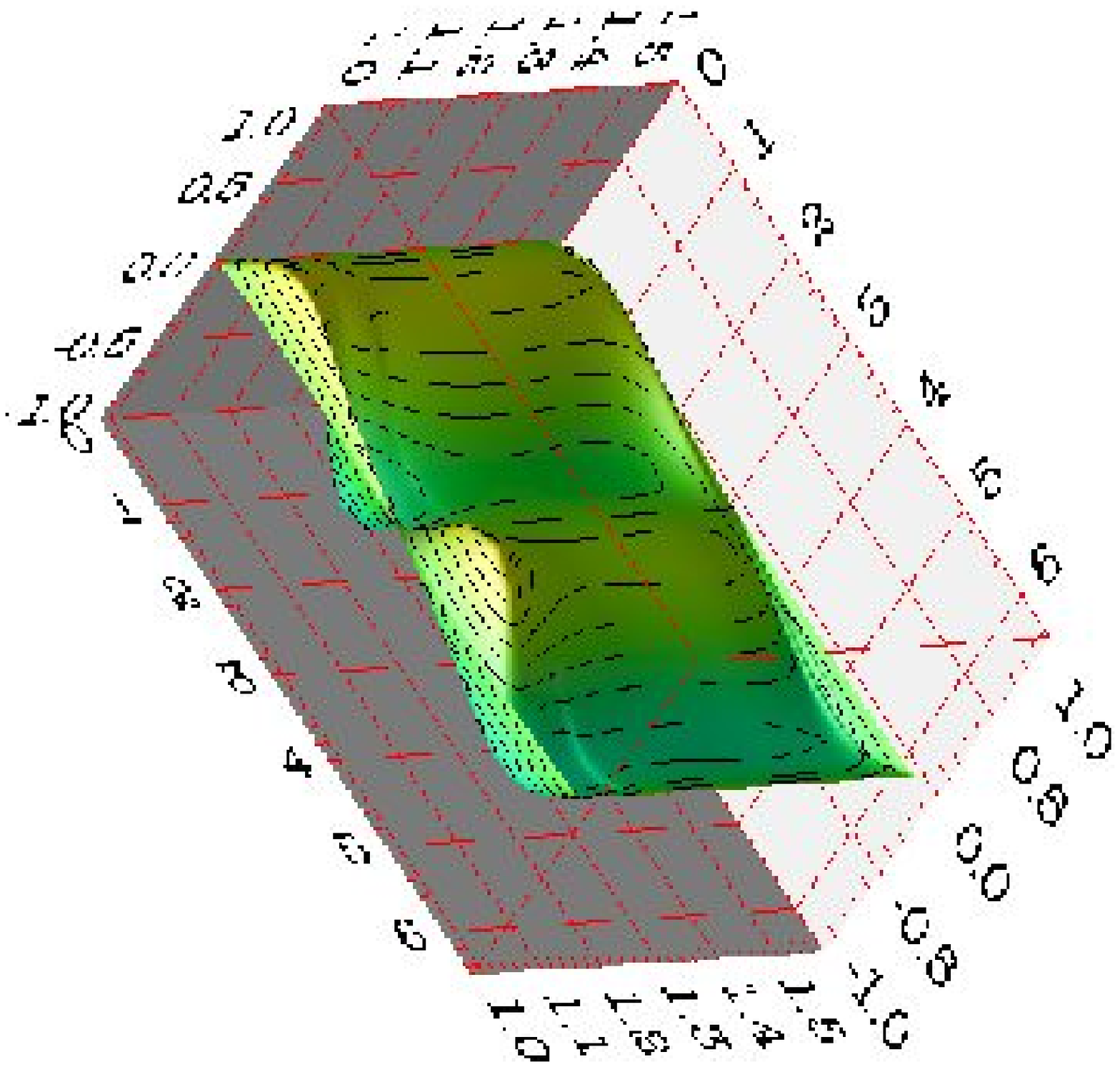}}
\end{figure}

\begin{figure}
\caption{The observable $P_x^\odot$, showing its sensitivity to the coupling constants $G_v^\phi$ and 
$G_t^\phi$. (a): $G_v^\phi=4,\,\, G_t^\phi=0$; (b): $G_v^\phi=-4,\,\, 
G_t^\phi=0$; (c): $G_v^\phi=0,\,\, G_t^\phi=4$; (d): $G_v^\phi=0,\,\, 
G_t^\phi=-4$. In each case, the surface is shown as a function of 
$m_{K\overline{K}}$ and $\Phi^*$.\label{phib}}
\vskip 0.2in
\centerline{(a)\includegraphics[height=3.0in,angle=0]{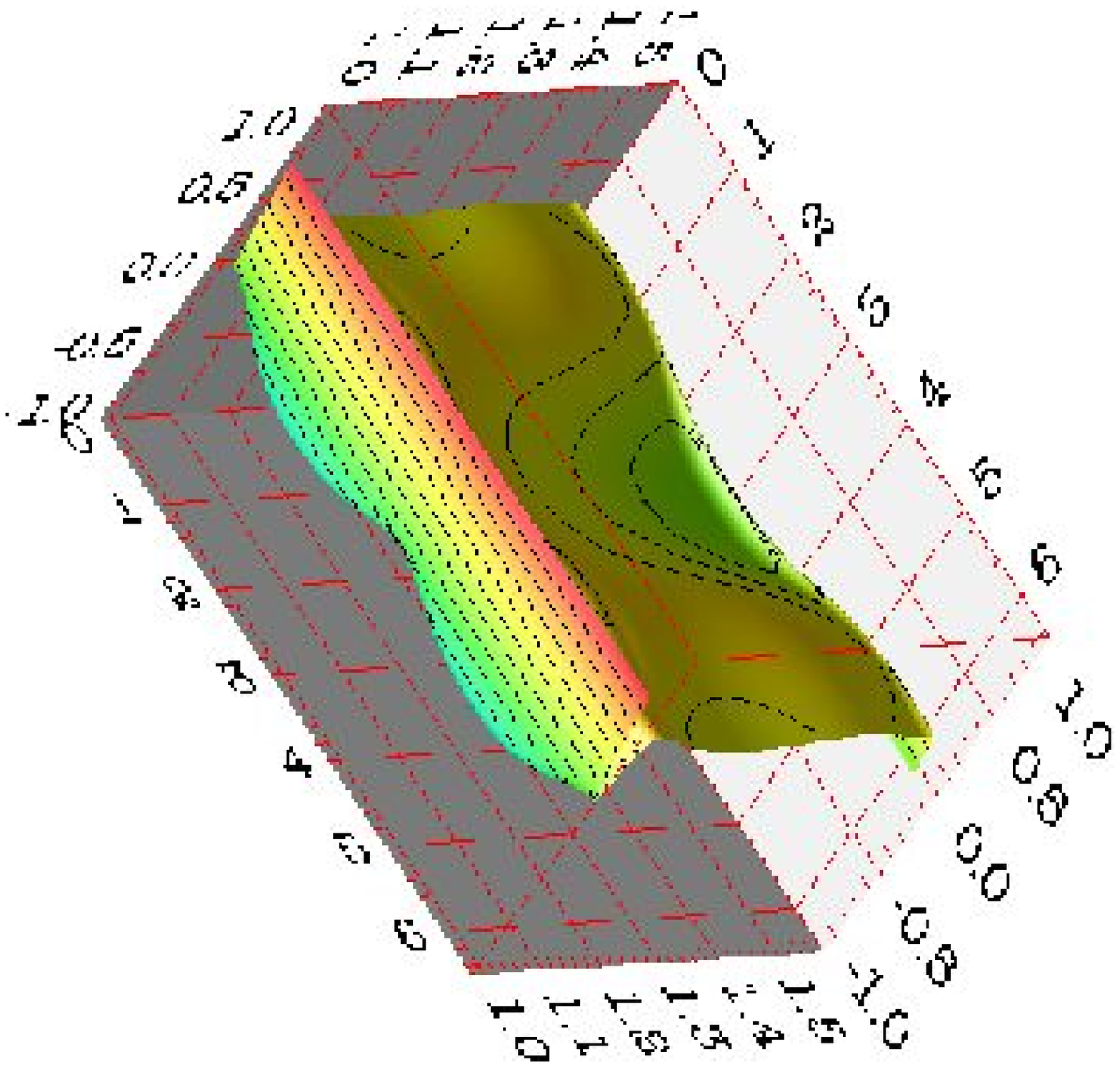}
(b)\includegraphics[height=3.0in,angle=0]{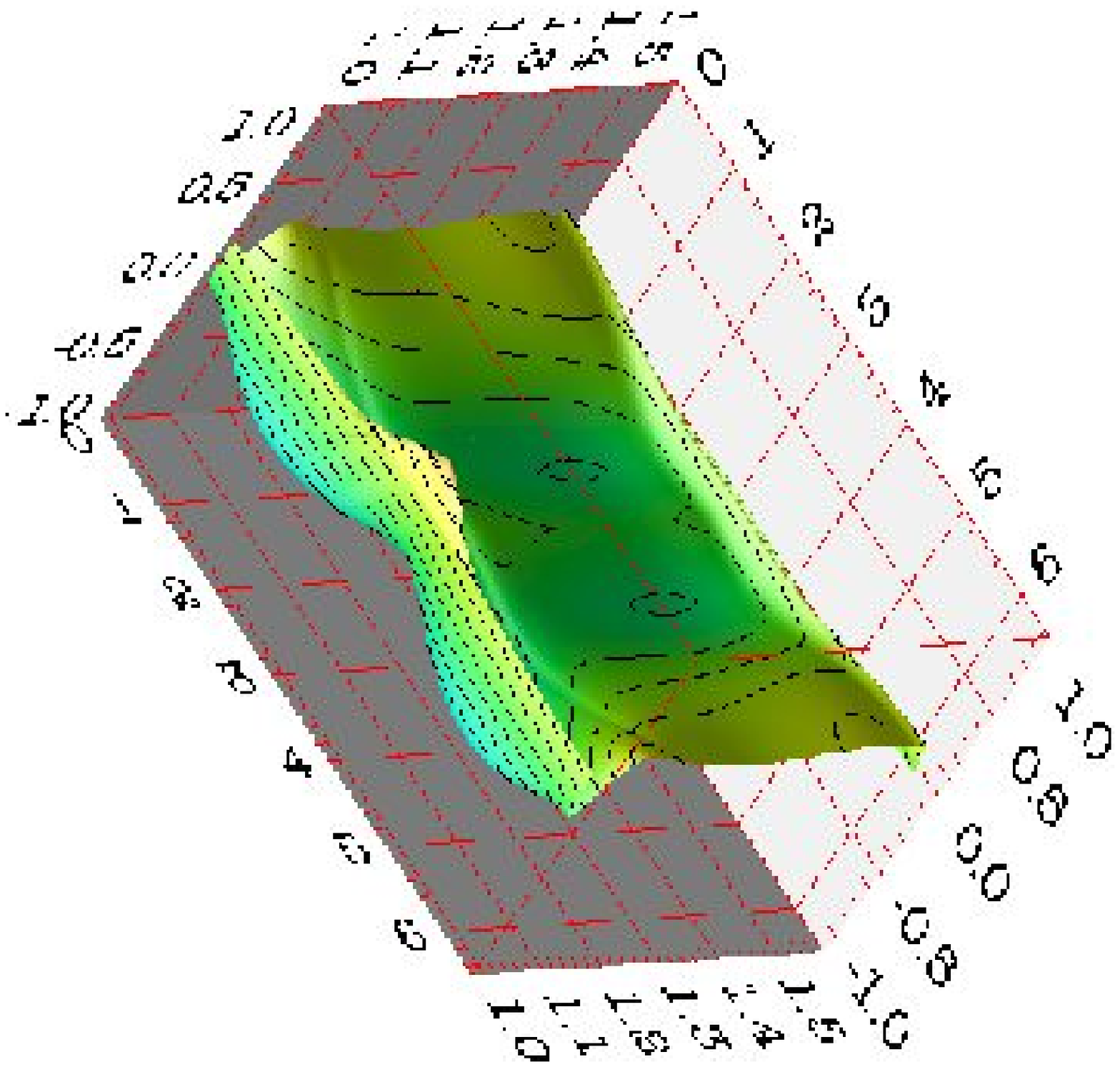}}
\vskip0.2in
\centerline{(c)\includegraphics[height=3.0in,angle=0]{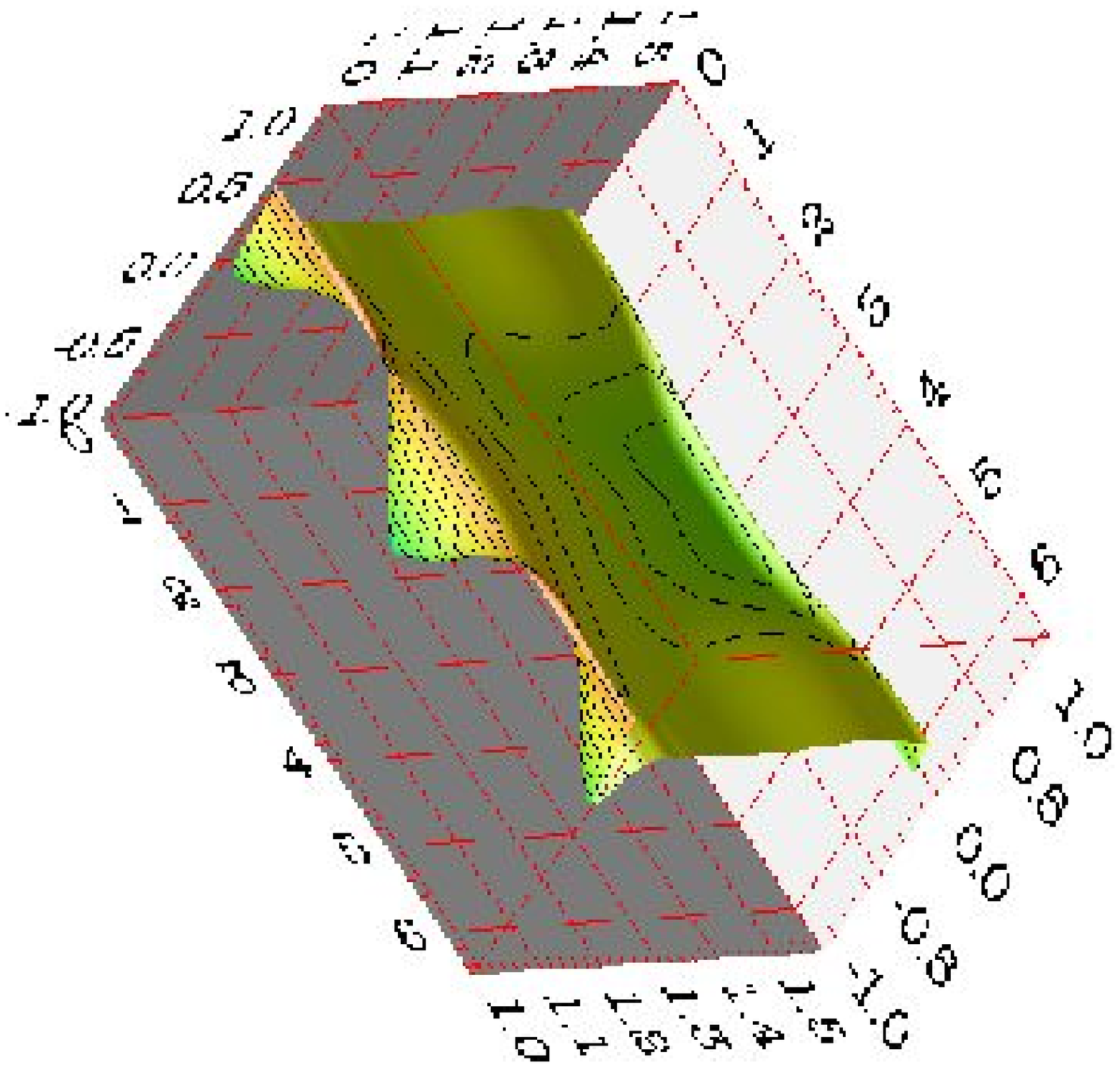}
(d)\includegraphics[height=3.0in,angle=0]{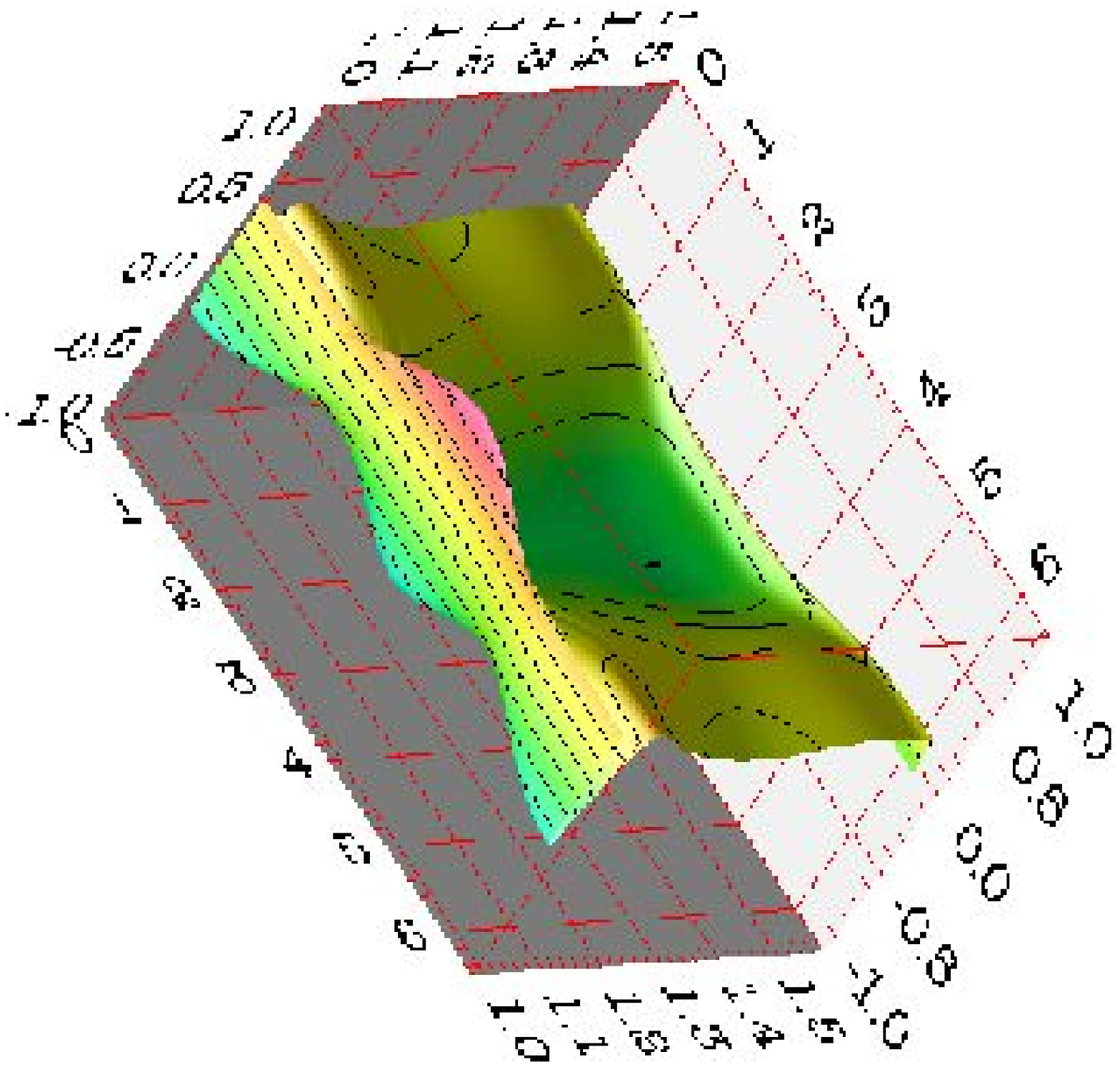}}
\end{figure}

The coupling of the $\phi$ meson to the nucleon is described by a
phenomenological Lagrangian that takes the form
\begin{equation}\label{equa98b_sing}
{\cal L}= \overline{N}\left(G_v^\phi \gamma^\mu \phi_\mu
+i\frac{G_t^\phi}{2 M_N} \gamma^\mu \gamma^\nu\left(\partial_\nu 
\phi_\mu\right)\right)N 
\end{equation}
The two constants $G_v^\phi$ and $G_t^\phi$ are not well known. Figures \ref{phia} and \ref{phib} show the
effects of choosing different values for these constants, on the observables $I^s$ (fig. \ref{phia}) and 
$P_x^\odot$ (fig. \ref{phib}). In each of these two figures, (a) corresponds to the choice 
$G_v^\phi=4,\,\, G_t^\phi=0$; (b) corresponds to $G_v^\phi=-4,\,\, 
G_t^\phi=0$; (c) corresponds to $G_v^\phi=0,\,\, G_t^\phi=4$; and (d) corresponds to $G_v^\phi=0,\,\, 
G_t^\phi=-4$. The process is $\gamma p\to pK^0\overline{K}^0$.

In all of the figures, the effects of the $\phi$ show up most clearly near its resonant mass of 1.02 MeV
(the second independent variable in these plots is $s_{K\overline{K}}$). Note the changes in the
observables as the coupling constants are changed. This indicates that polarization observables can be
used to help pin down coupling constants such as the ones explored here.

\subsection{Sub-threshold Resonance}

\begin{figure}
\caption{The observable $P_x^s$, showing its sensitivity to the sub-threshold resonance $\Lambda(1405)$.
(a) results with the $\Lambda(1405)$ included in the calculation. (b) results
when this state is excluded. Both surfaces are shown as functions of 
$m_{N\overline{K}}$ and $\Phi^*$.\label{lambdaa}}
\vskip 0.2in
\centerline{(a)\includegraphics[height=4.0in,angle=0]{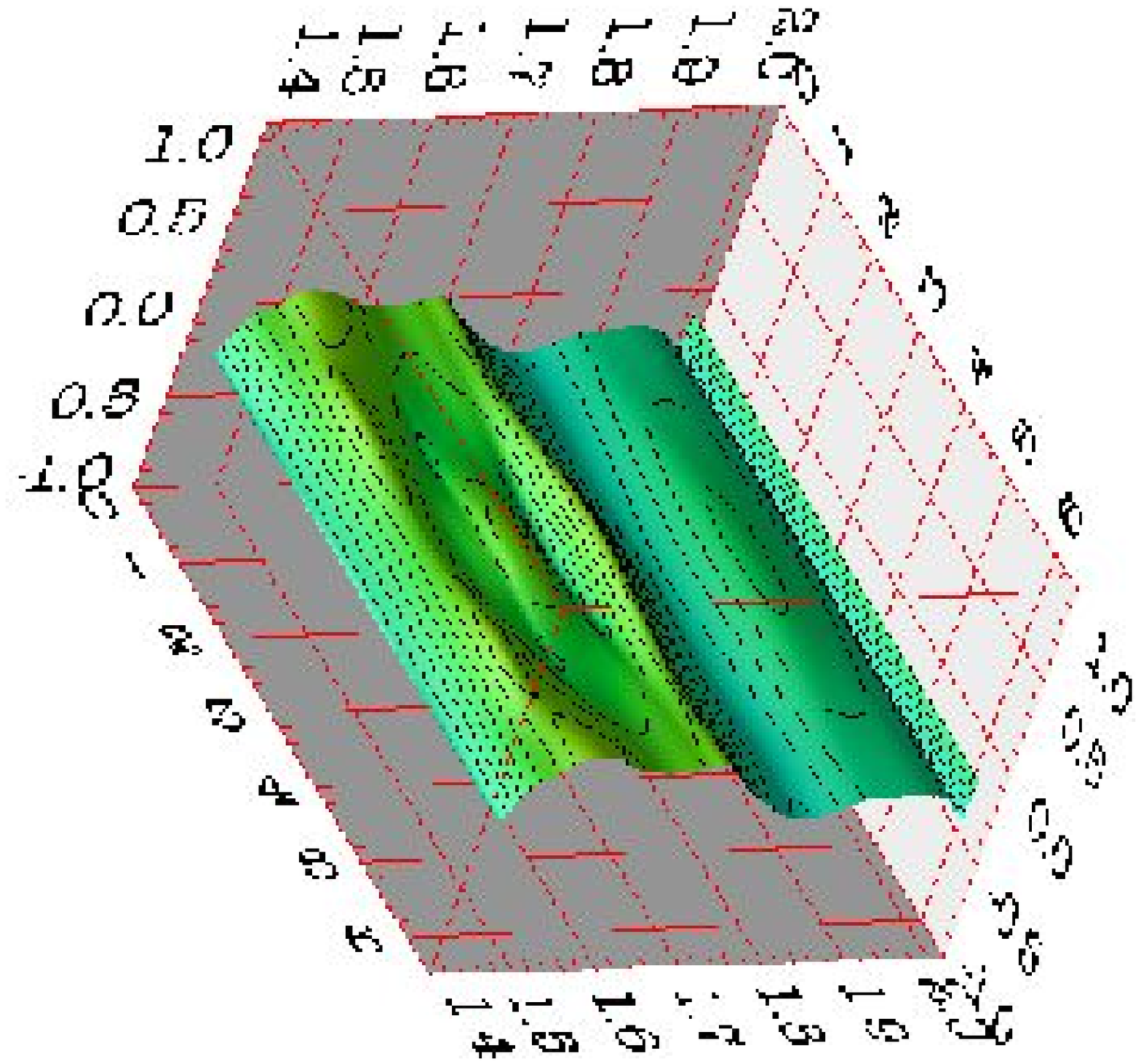}}
\vskip0.2in
\centerline{(b)\includegraphics[height=4.0in,angle=0]{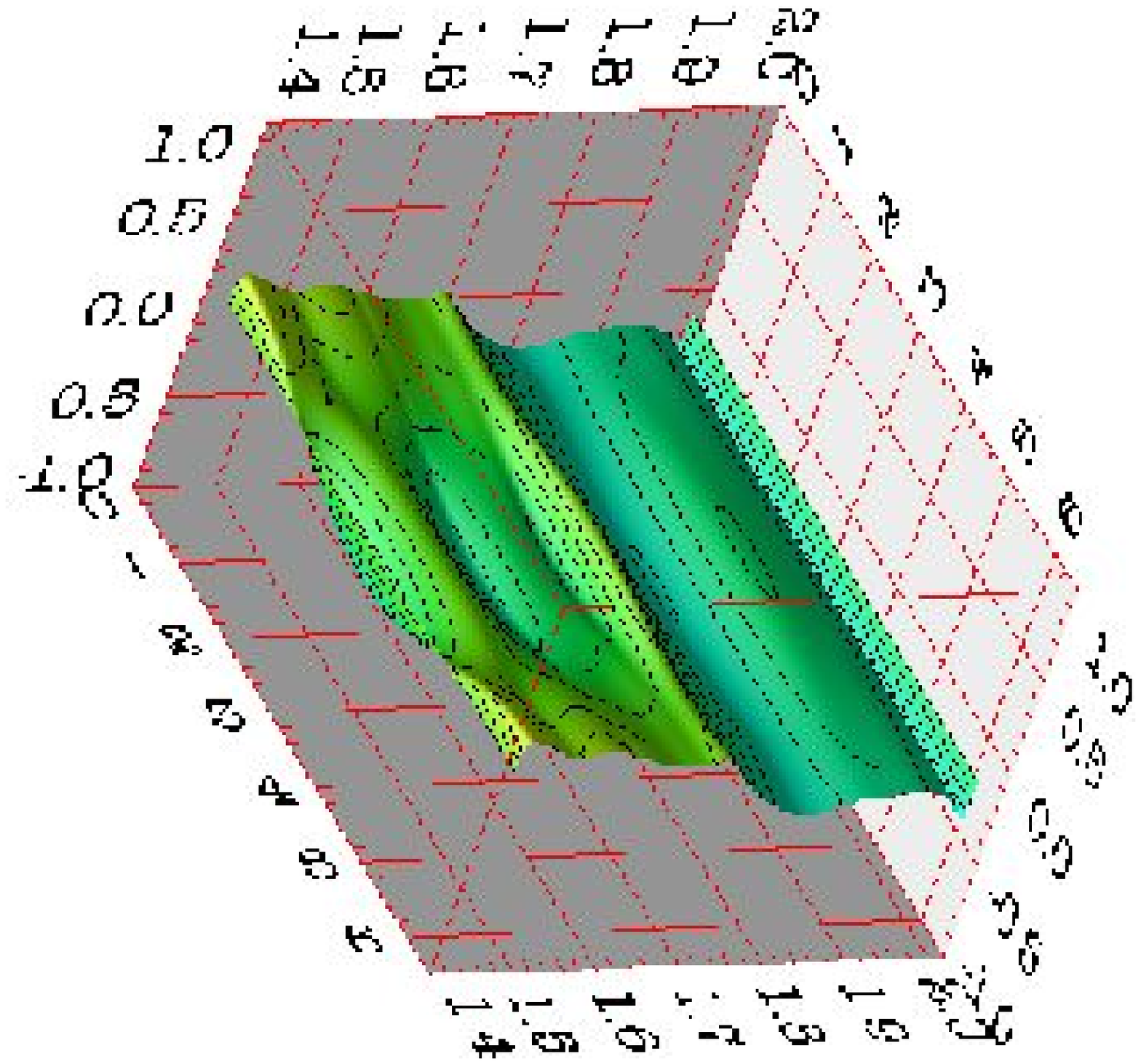}}
\end{figure}

\begin{figure}
\caption{The observable $I^c$, showing its sensitivity to the sub-threshold resonance $\Lambda(1405)$.
(a) results with the $\Lambda(1405)$ included in the calculation. (b) results
when this state is excluded. Both surfaces are shown as functions of 
$m_{N\overline{K}}$ and $\Phi^*$.\label{lambdab}}
\vskip 0.2in
\centerline{(a)\includegraphics[height=4.0in,angle=0]{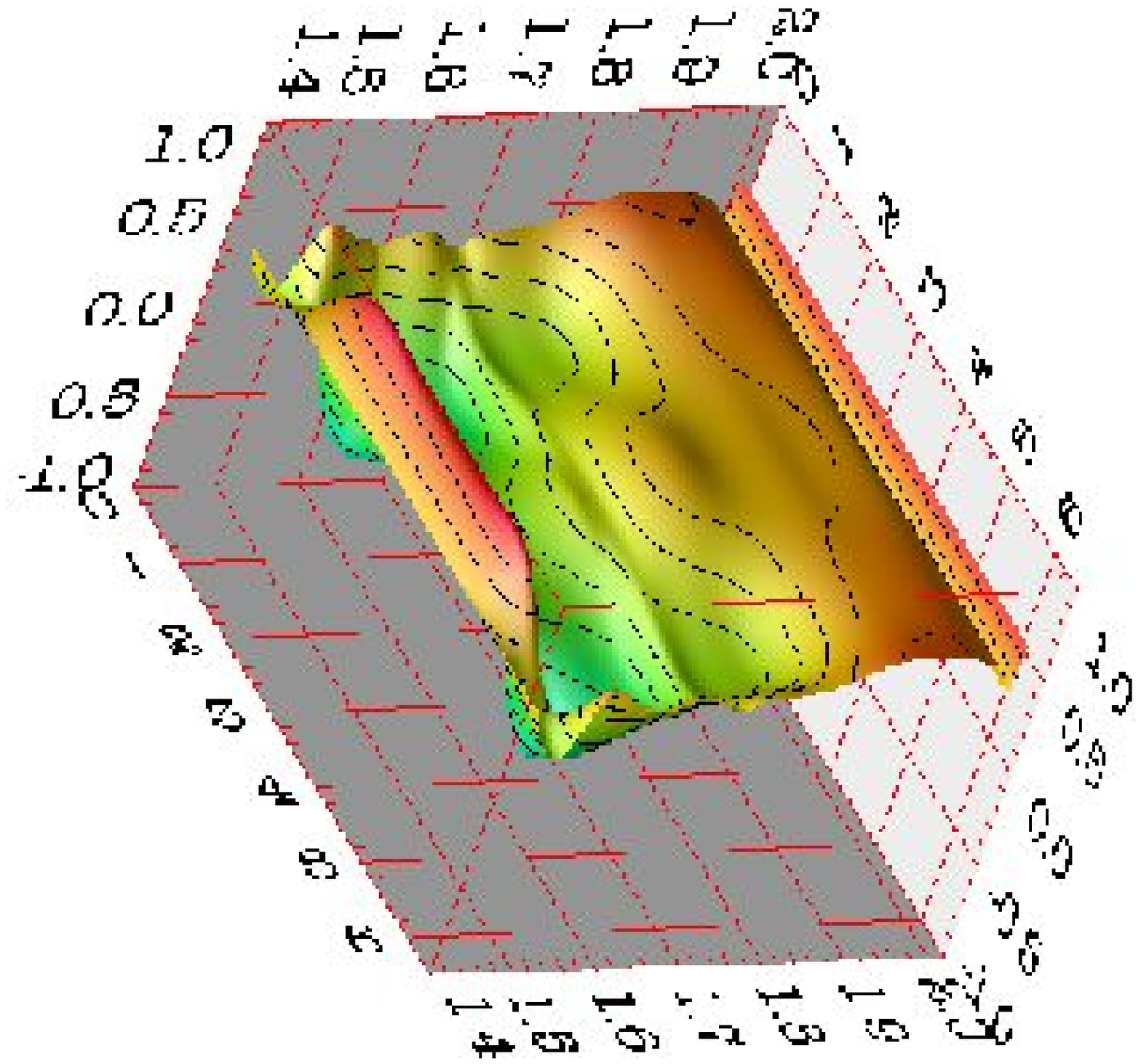}}
\vskip0.2in
\centerline{(b)\includegraphics[height=4.0in,angle=0]{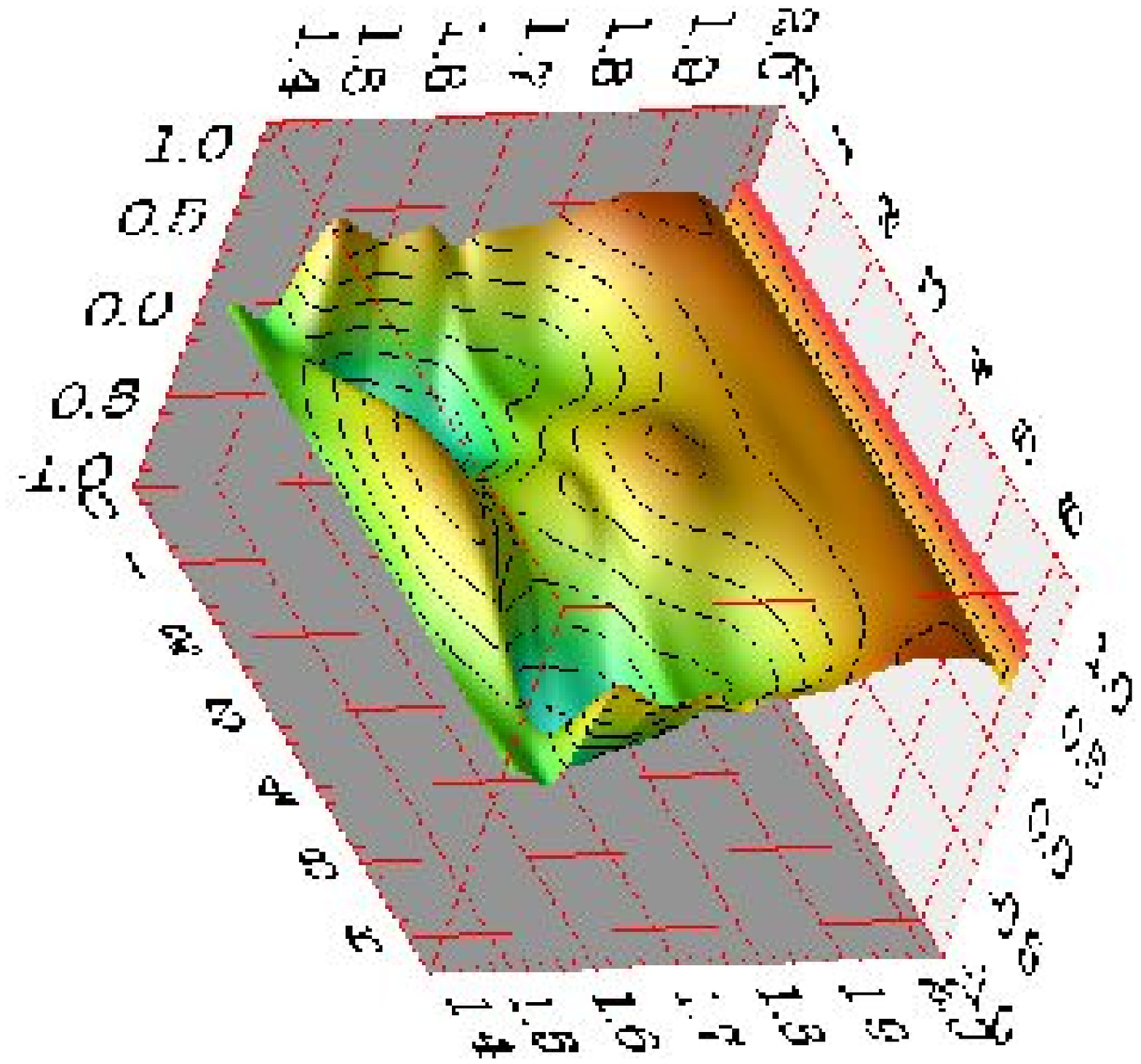}}
\end{figure}

\begin{figure}
\caption{The observable $P_y^c$, showing its sensitivity to the sub-threshold resonance $\Lambda(1405)$.
(a) results with the $\Lambda(1405)$ included in the calculation. (b) results
when this state is excluded. Both surfaces are shown as functions of 
$m_{N\overline{K}}$ and $\Phi^*$.\label{lambdac}}
\vskip 0.2in
\centerline{(a)\includegraphics[height=4.0in,angle=0]{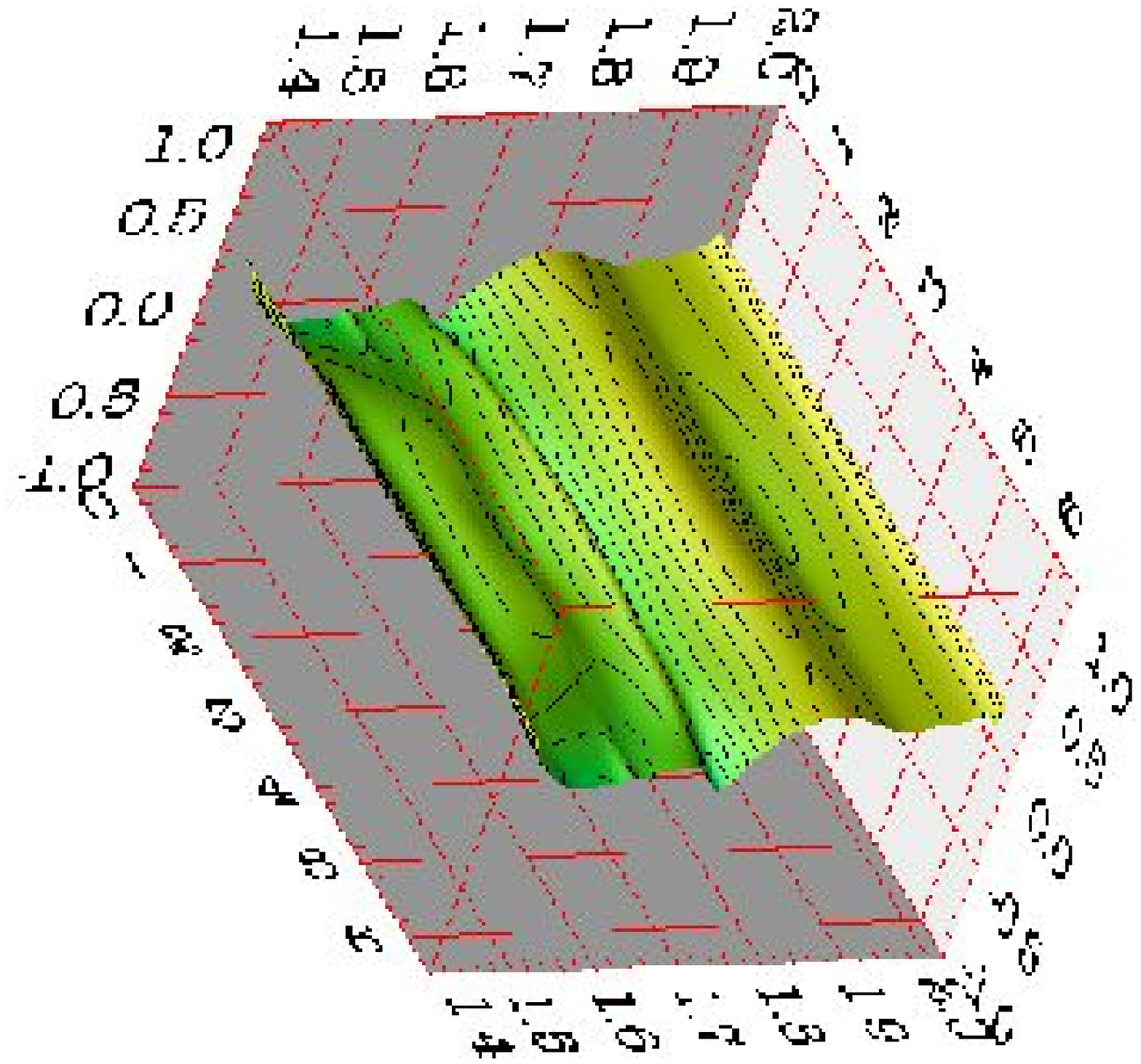}}
\vskip0.2in
\centerline{(b)\includegraphics[height=4.0in,angle=0]{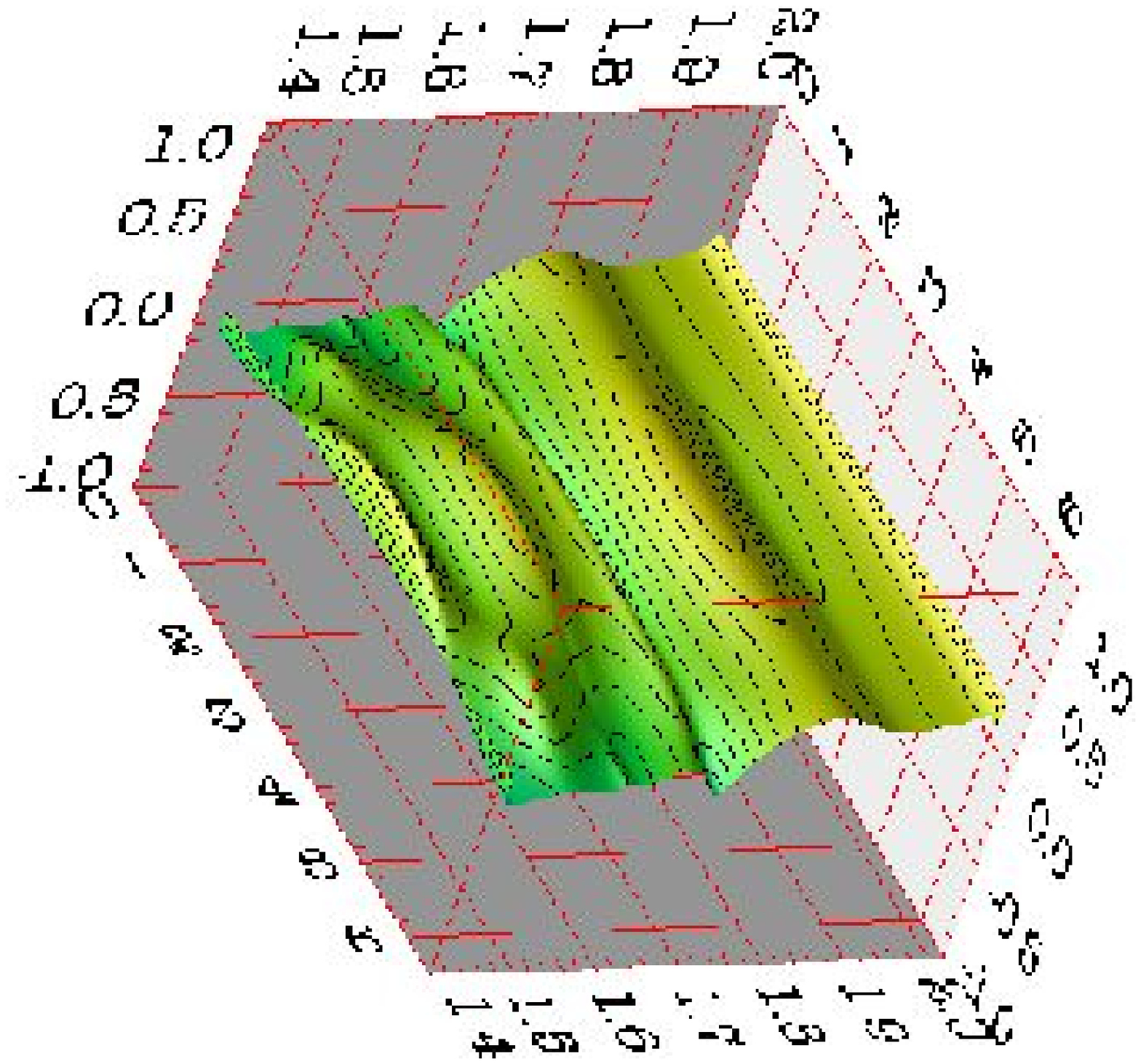}}
\end{figure}

\begin{figure}
\caption{Same as in fig. \ref{lambdac}, but with a different view point.\label{lambdacp}}
\vskip 0.2in
\centerline{(a)\includegraphics[height=4.0in,angle=270]{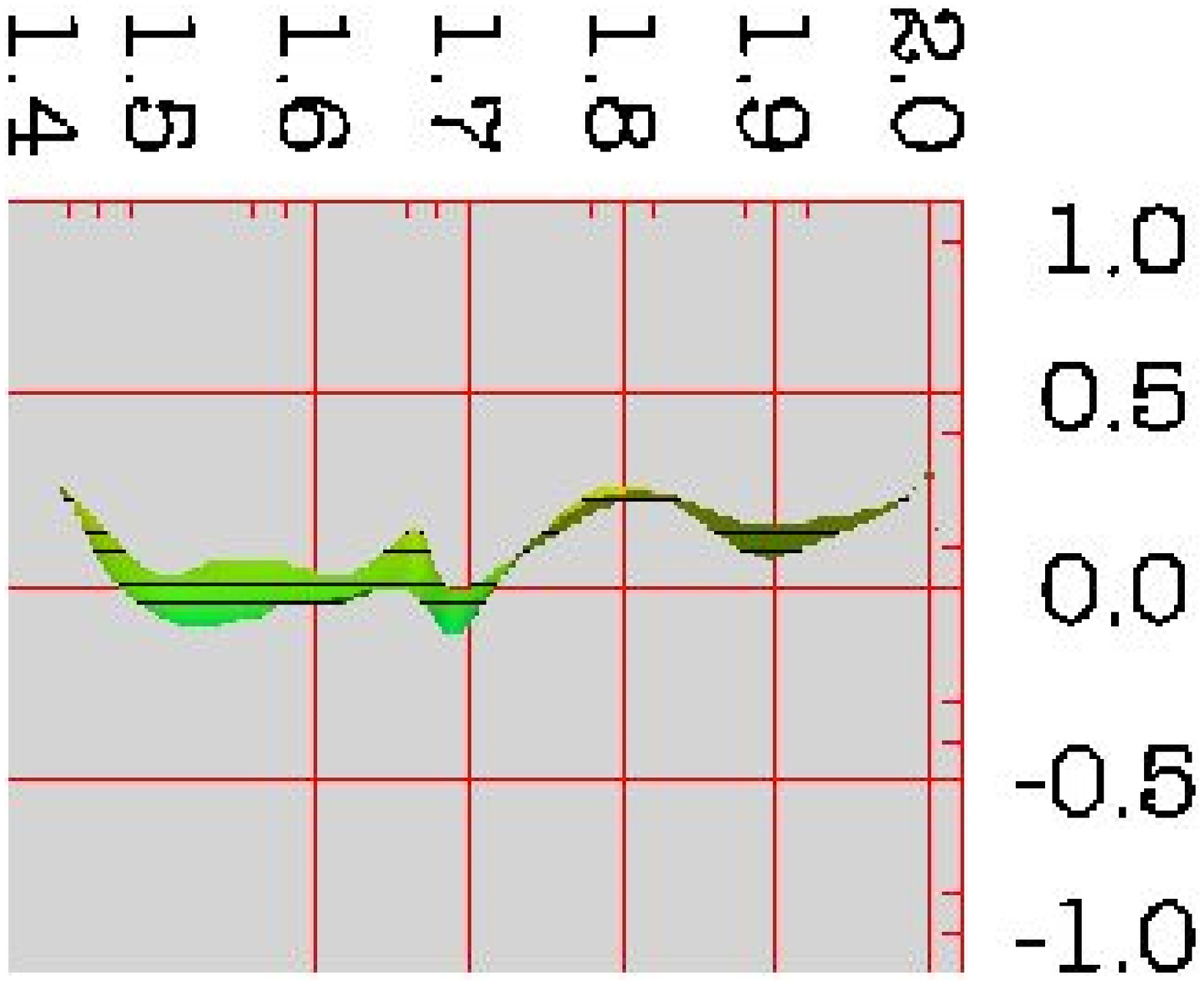}}
\vskip0.2in
\centerline{(b)\includegraphics[height=4.0in,angle=270]{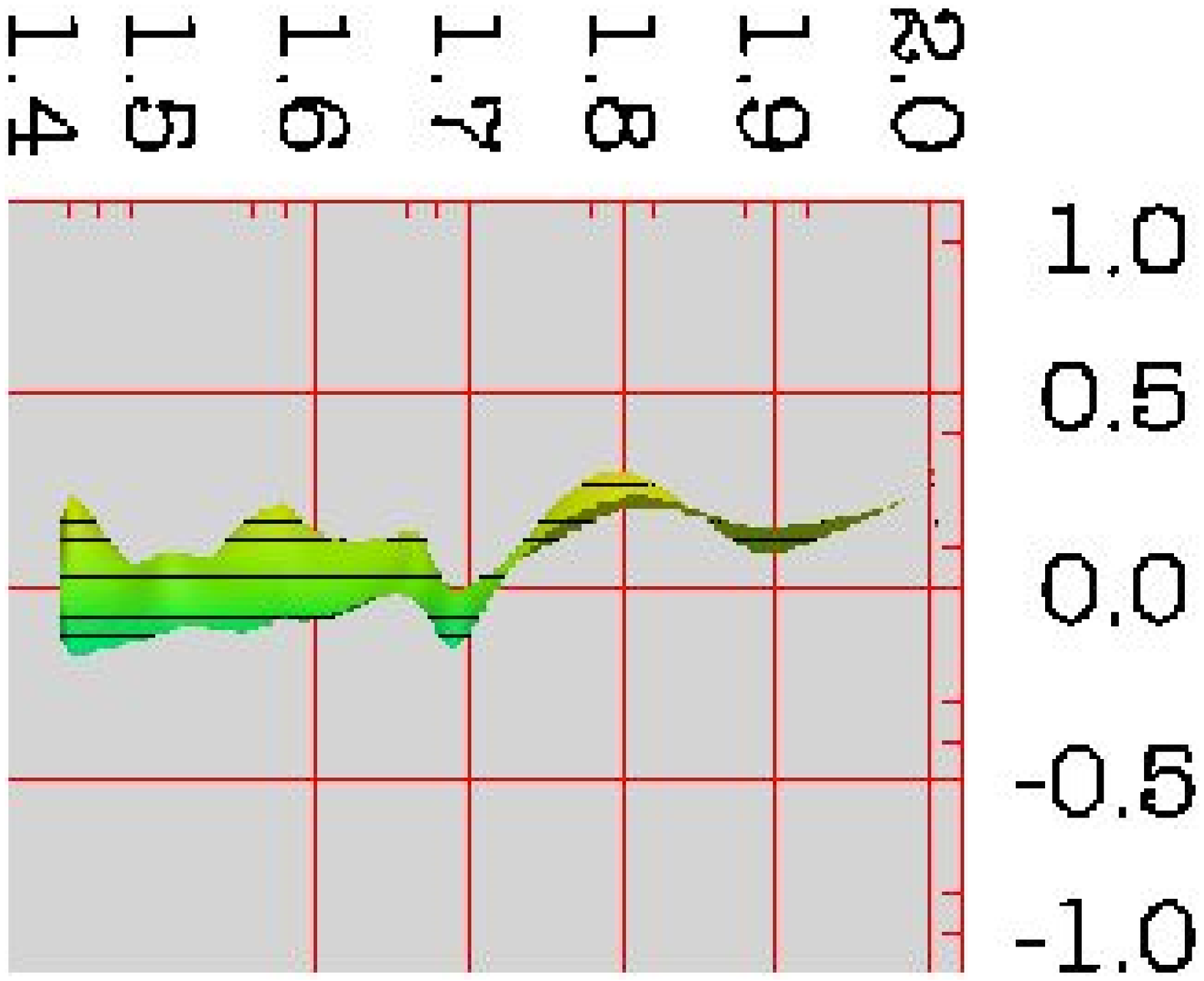}}
\end{figure}

\begin{figure}
\caption{The observable ${\cal O}_{yz^\prime}^c$, showing its sensitivity to the sub-threshold resonance $\Lambda(1405)$.
(a) results with the $\Lambda(1405)$ included in the calculation. (b) results
when this state is excluded. Both surfaces are shown as functions of 
$m_{N\overline{K}}$ and $\Phi^*$.\label{lambdad}}
\vskip 0.2in
\centerline{(a)\includegraphics[height=4.0in,angle=0]{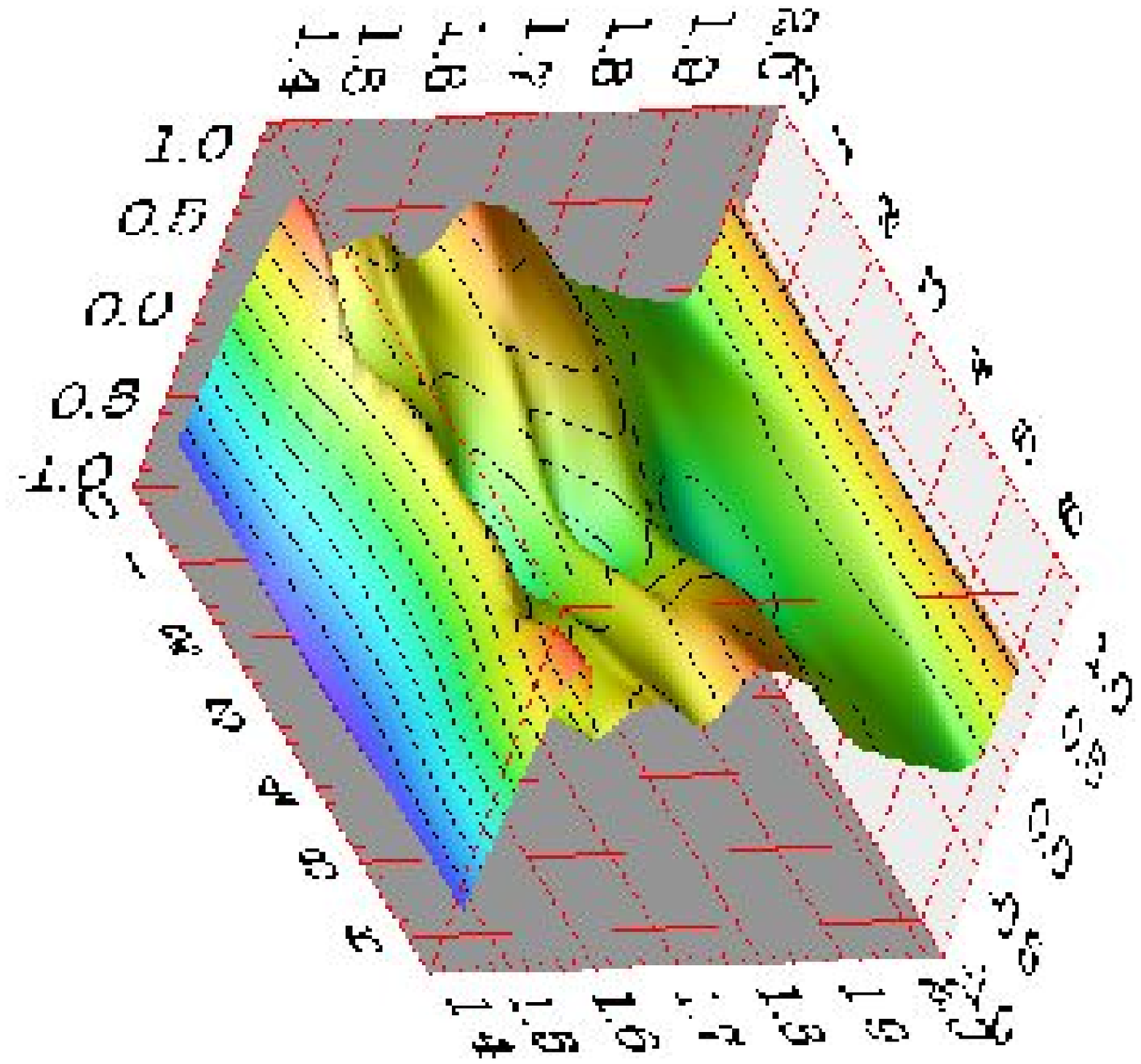}}
\vskip0.2in
\centerline{(b)\includegraphics[height=4.0in,angle=0]{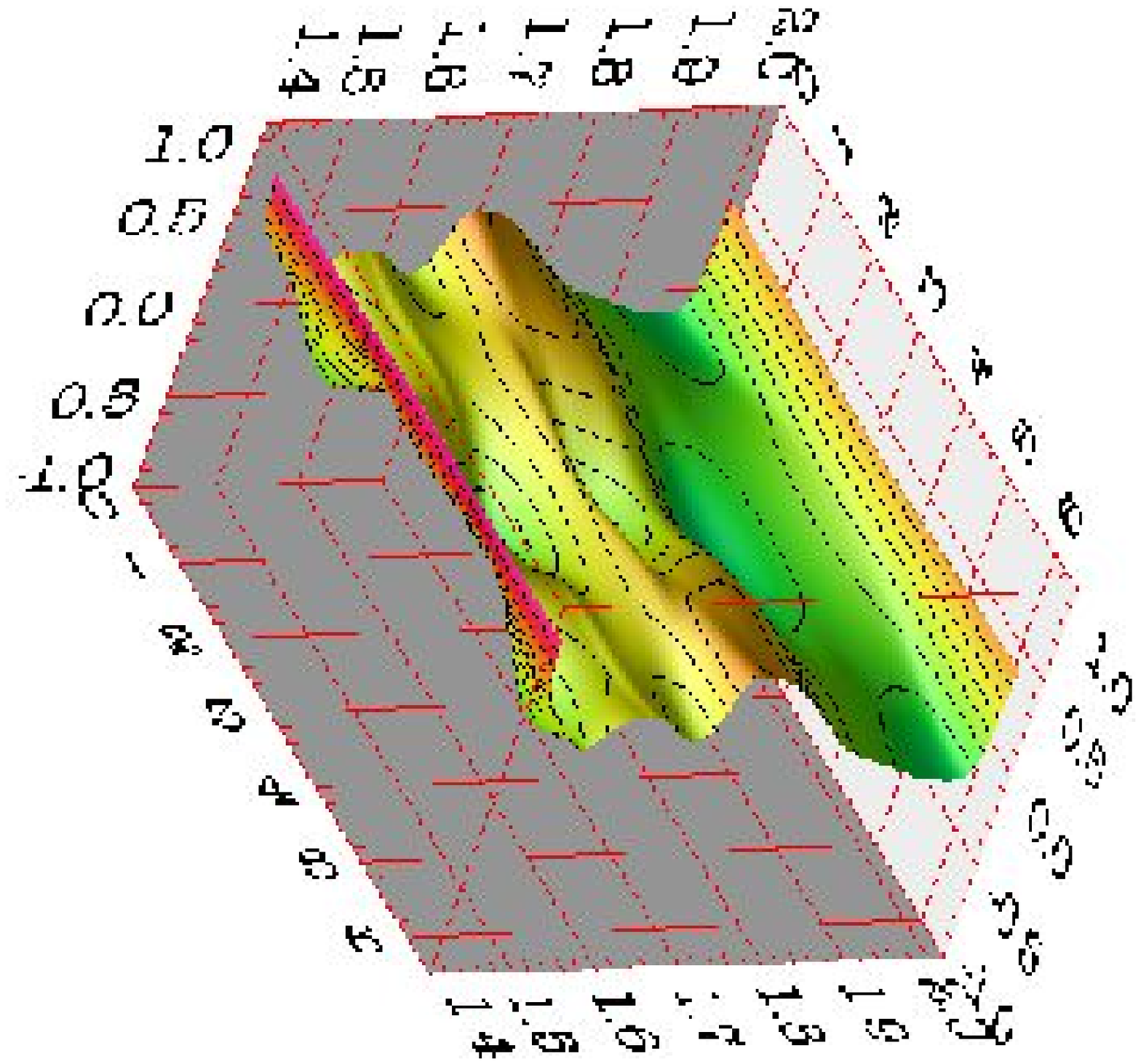}}
\end{figure}

The $\Lambda(1405)$ resonance is one of the relatively well-established hyperons. It lies just
below the $NK$ threshold, so its coupling (to $NK$) is not very well established. In the model
we use, we explore the sensitivity to this state by showing a few observables with this state
included in the calculation, and the same observables with the state excluded. The observables
we examine are $P_x^s$ (fig. \ref{lambdaa}), $I^c$ (fig. \ref{lambdab}), $P_y^c$ (figs.
\ref{lambdac} and \ref{lambdacp}), and ${\cal O}_{yz^\prime}^c$ (fig. \ref{lambdad}), in the
process $\gamma p\to nK^+\overline{K}^0$.

In all of these figures, the surfaces with the $\Lambda(1405)$ are very
different from those without it, especially near the lowest values of
$m_{N\overline{K}}$. Note, however, that the effects of the state can be seen
fairly far away from threshold, especially in fig. \ref{lambdac}. At values of
$\Phi^*$ near $\pi$ (near the middle of the $\Phi^*$ axis), and for 
$m_{K\overline{K}}$ near 1.6 GeV (about 200 MeV away from the nominal mass of
the $\Lambda(1405)$, there is a local minimum in the observable when the
$\Lambda(1405)$ is included in the calculation, and a local maximum when it is
excluded. This is best seen by looking at the two surfaces `edge-on', as shown
in fig. \ref{lambdacp}. Then the differences that arise, even 200 MeV or more
away from the nominal mass of the $\Lambda(1405)$, are clearly visible. This
illustrates that, in calculations such as this, `small' contributions may not
affect the cross section much, but can have significant effects in polarization
observables, even in regions where one might expect their effects to be small.

\subsection{Pentaquark Search}

One very interesting question regarding these observables is their possible sensitivity to exotic resonances,
such as the $\Theta^+$ \cite{penta}. If observables are found that show sensitivity to this state, they can be used
to confirm its existence (or otherwise), assuming production mechanisms like those presented in
\cite{zplus}. One of the disadvantages of using the differential cross section to search for states 
like this is the fact that one state (or a few states) may provide a very large
background against which a small signal must be sought. With polarization observables, this is not
necessarily the case, and the surfaces that we show illustrate what might be possible in pentaquark
searches.

\begin{figure}
\caption{The beam asymmetry, $I^\odot$, showing its sensitivity to the exotic resonance $\Theta^+$.
(a) results with the $\Theta^+$ included in the calculation. (b) results
when this state is excluded. Both surfaces are shown as functions of 
$m_{NK}$ and $\Phi^*$.\label{thetaa}}
\vskip 0.2in
\centerline{(a)\includegraphics[height=4.0in,angle=0]{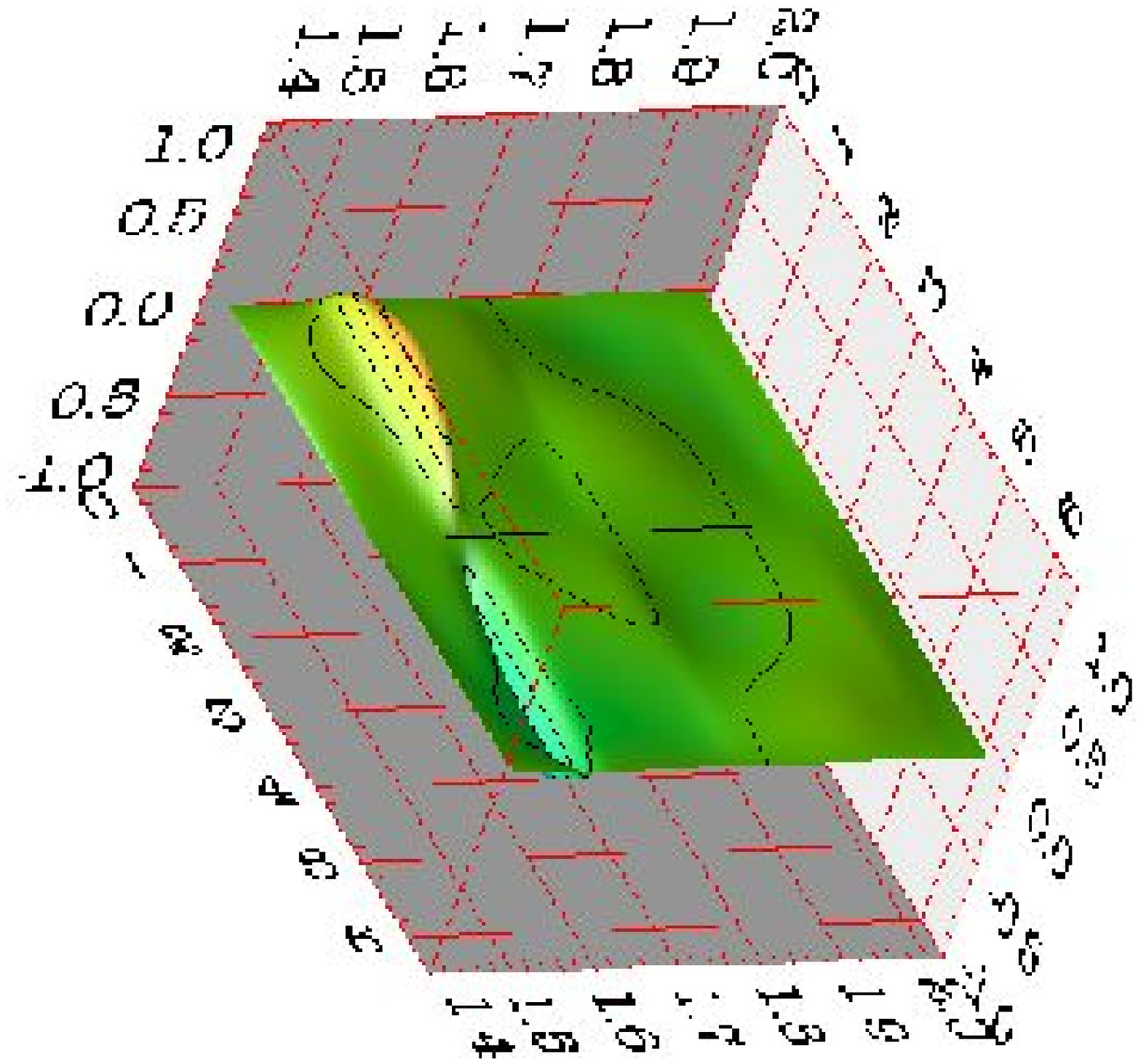}}
\vskip0.2in
\centerline{(b)\includegraphics[height=4.0in,angle=0]{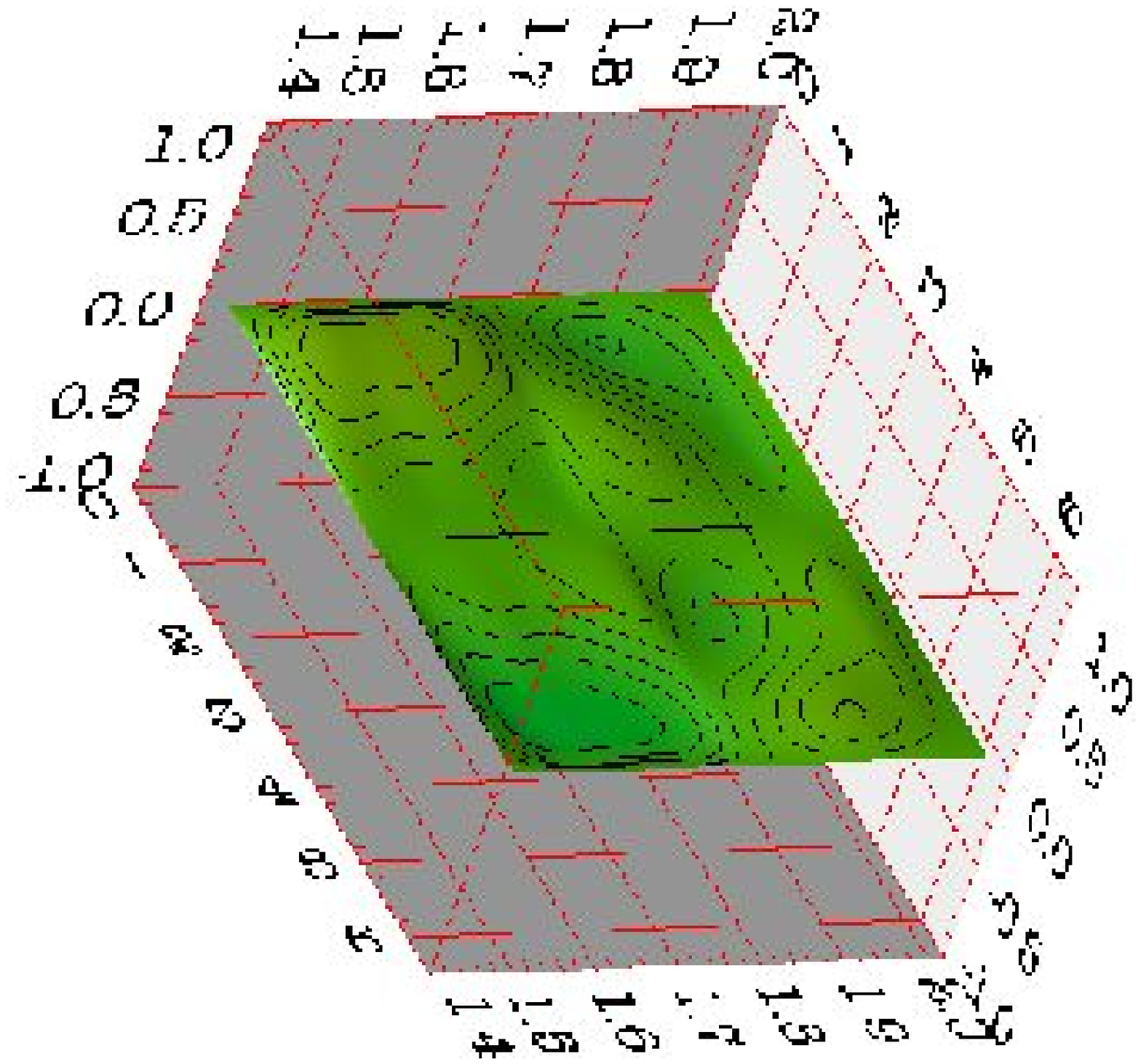}}
\end{figure}

\begin{figure}
\caption{The observable $P^\odot_x$, showing its sensitivity to the exotic resonance $\Theta^+$.
(a) results with the $\Theta^+$ included in the calculation. (b) results
when this state is excluded. Both surfaces are shown as functions of 
$m_{NK}$ and $\Phi^*$.\label{thetab}}
\vskip 0.2in
\centerline{(a)\includegraphics[height=4.0in,angle=0]{pol_005_spp1_17.ps}}
\vskip0.2in
\centerline{(b)\includegraphics[height=4.0in,angle=0]{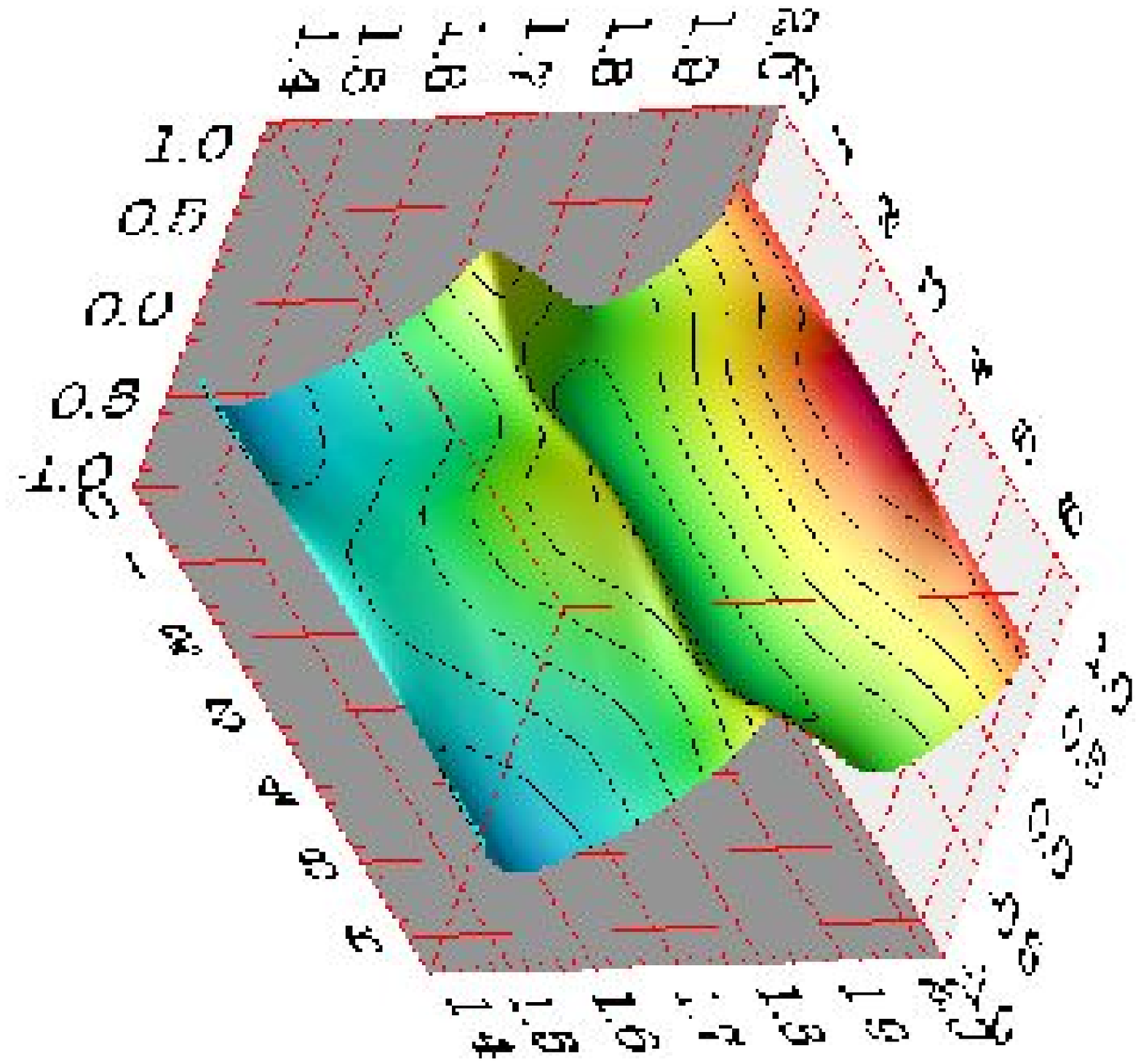}}
\end{figure}

\begin{figure}
\caption{The observable $P^\odot_z$, showing its sensitivity to the exotic resonance $\Theta^+$.
(a) results with the $\Theta^+$ included in the calculation. (b) results
when this state is excluded. Both surfaces are shown as functions of 
$m_{NK}$ and $\Phi^*$.\label{thetac}}
\vskip 0.2in
\centerline{(a)\includegraphics[height=4.0in,angle=0]{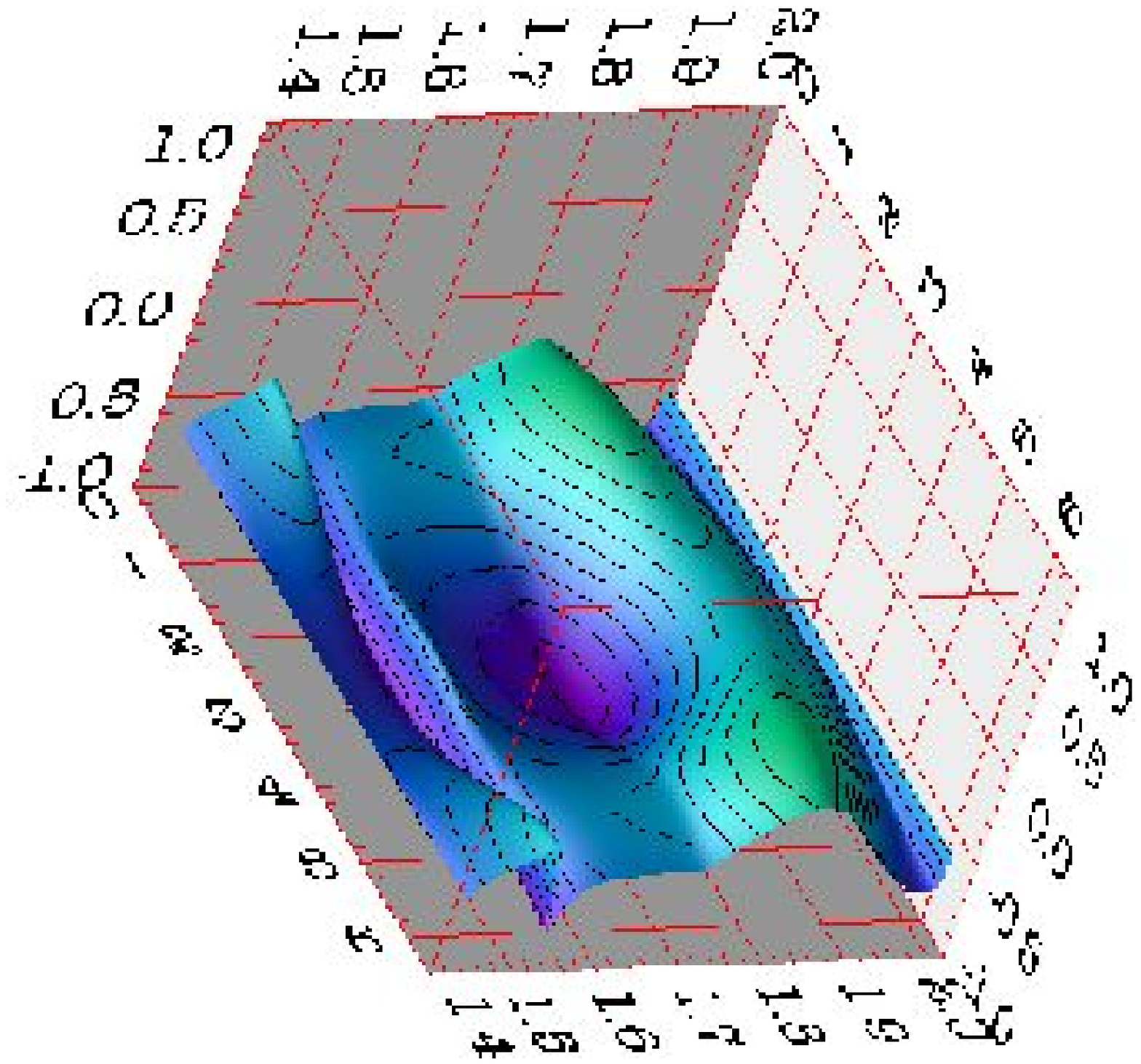}}
\vskip0.2in
\centerline{\includegraphics[height=4.0in,angle=0]{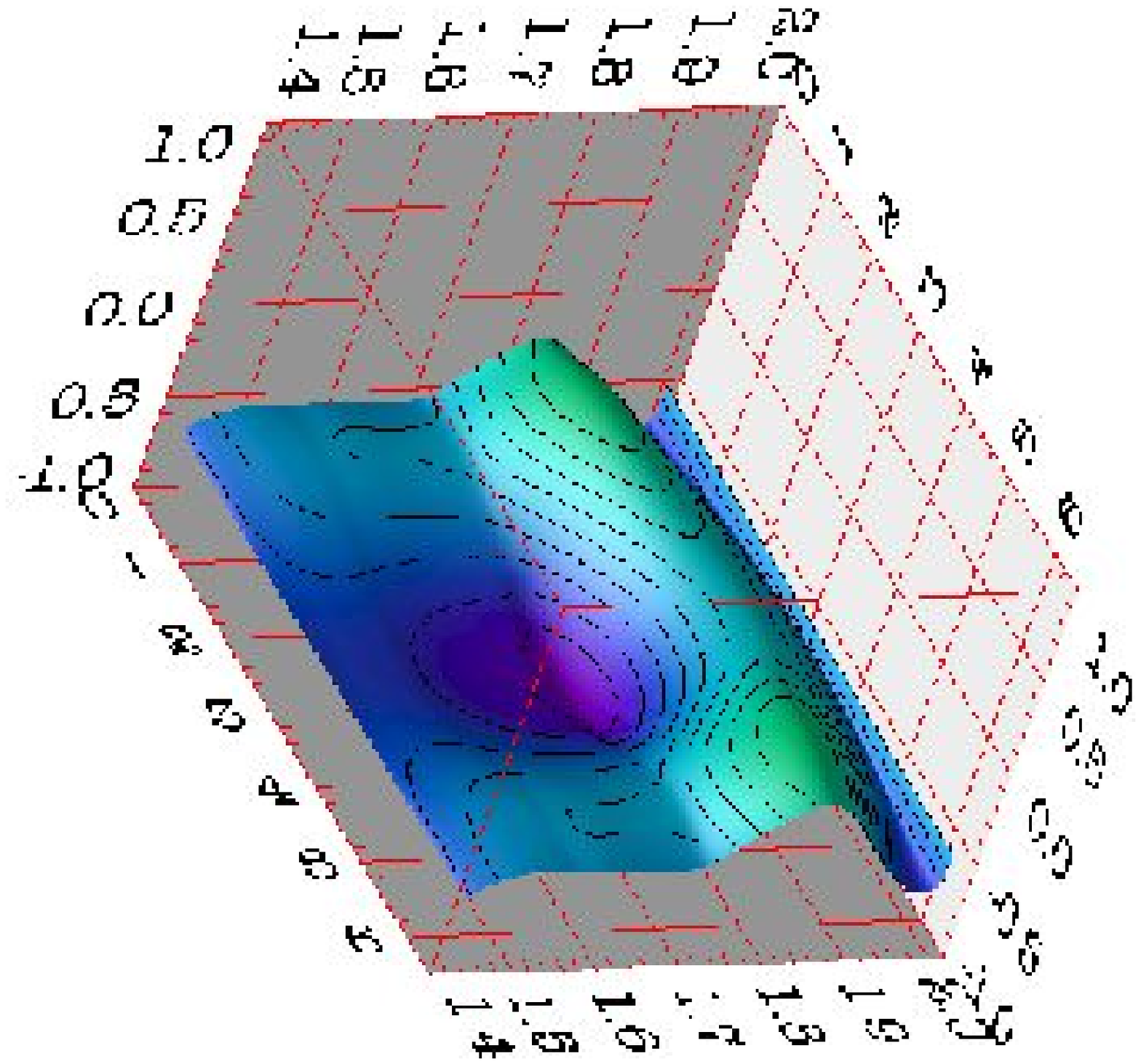}}
\end{figure}

Figs. \ref{thetaa} to \ref{thetac} ($I^\odot$, $P_x^\odot$ and $P_z^\odot$,
respectively) show the surfaces that result when a $\Theta^+$ with $J^P=1/2^+$
is included in the calculation (the surfaces in (a)), and when it is excluded
(b), for the process $\gamma p\to nK^+\overline{K}^0$. In fig. \ref{thetaa},
the helicity asymmetry without the pentaquark has an absolute maximum value
that is less than 0.1. With the pentaquark, this observable ranges between -0.5
and +0.5, but only in the immediate vicinity of the pentaquark's mass (in the
$nK^+$ system). Note that this asymmetry has already been measured at JLab for
$\gamma p\to p\pi^+\pi^-$ \cite{strauch}. Thus, it may be
possible to measure it for $\gamma N\to NK\overline{K}$ relatively quickly.

Figs. \ref{thetab} and \ref{thetac} show similar structures in the surfaces
$P_x^\odot$ and  $P_z^\odot$. Note that in all cases, the structures stand out
clearly for two reasons. The first is that the pentaquark is a narrow state (in
this calculation, the width is 10 MeV). The `width' of any structure that might
be observed will be similar to that of the state giving rise to the structure.
The second reason is that the $\Theta^+$ is the {\it only} resonance in the
$nK^+$ channel. All other resonances are in the $n\overline{K}^0$ channel. The
kinematic reflections of these resonances will show up, as can be seen in figs
\ref{thetab} and \ref{thetac}, but the presence of the $\Theta^+$ in this
channel has a marked effect. Note that in fig. \ref{thetac}, the predominantly
blue color means that this observable is predicted to be large (in the framework
of the model used) and negative, with values approaching -1 in some regions of
the surface.

\begin{figure}
\caption{The beam asymmetry, $I^\odot$, showing its sensitivity to the parity of the exotic resonance 
$\Theta^+$. (a) results when a $\Theta^+$ having $J^P=1/2^+$ is included. (b) 
results when this state has $J^P=1/2^-$. Both surfaces are shown as 
functions of $m_{NK}$ and $\Phi^*$.\label{thetad}}
\vskip 0.2in
\centerline{(a)\includegraphics[height=4.0in,angle=0]{pol_005_spp1_16.ps}}
\vskip0.2in
\centerline{(b)\includegraphics[height=4.0in,angle=0]{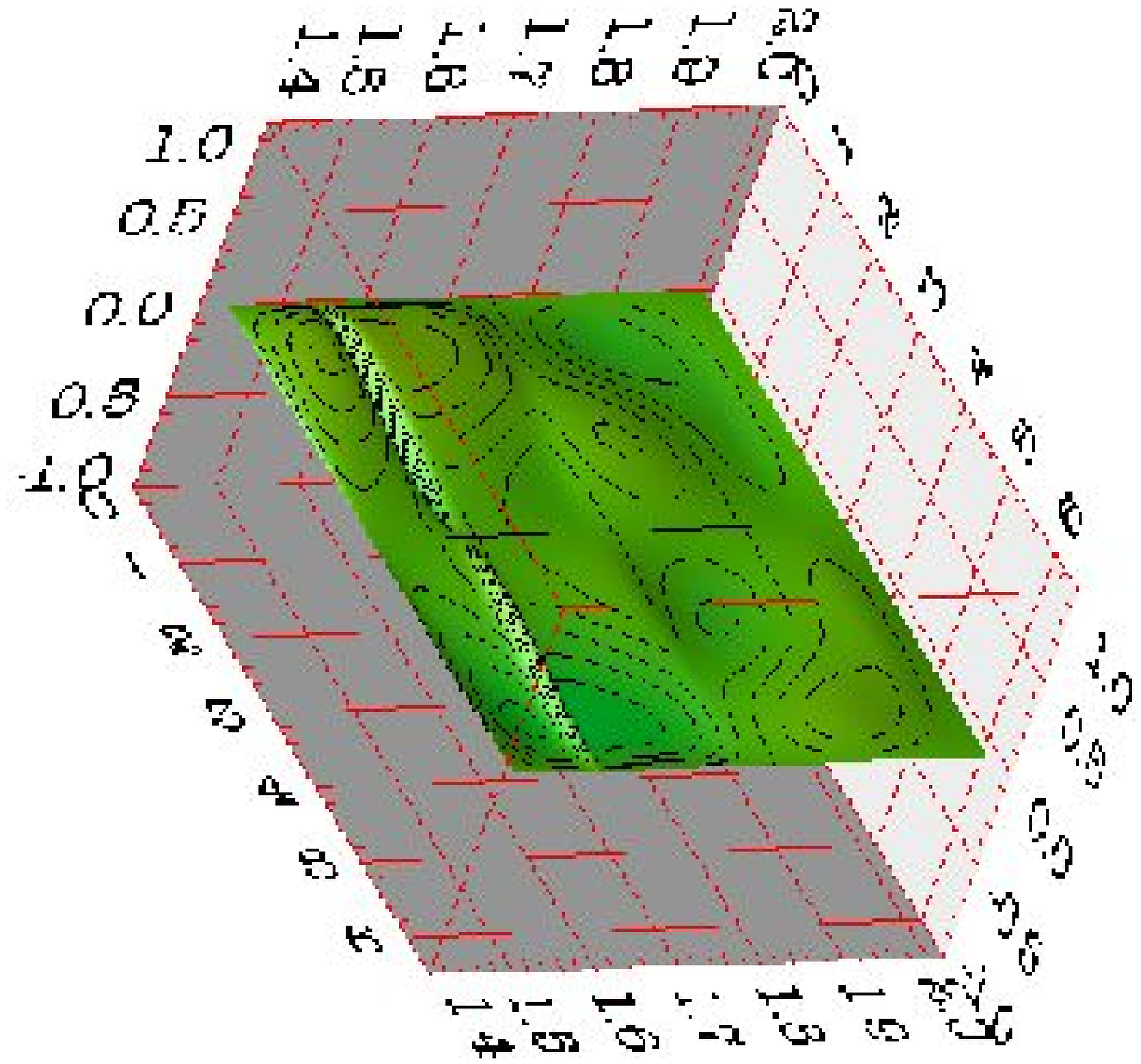}}
\end{figure}

\begin{figure}
\caption{The observable $P^\odot_x$, showing its sensitivity to the parity of the exotic resonance 
$\Theta^+$. (a) results when a $\Theta^+$ having $J^P=1/2^+$ is included. (b) 
results when this state has $J^P=1/2^-$. Both surfaces are shown as 
functions of $m_{NK}$ and $\Phi^*$.\label{thetae}}
\vskip 0.2in
\centerline{(a)\includegraphics[height=4.0in,angle=0]{pol_005_spp1_17.ps}}
\vskip0.2in
\centerline{(b)\includegraphics[height=4.0in,angle=0]{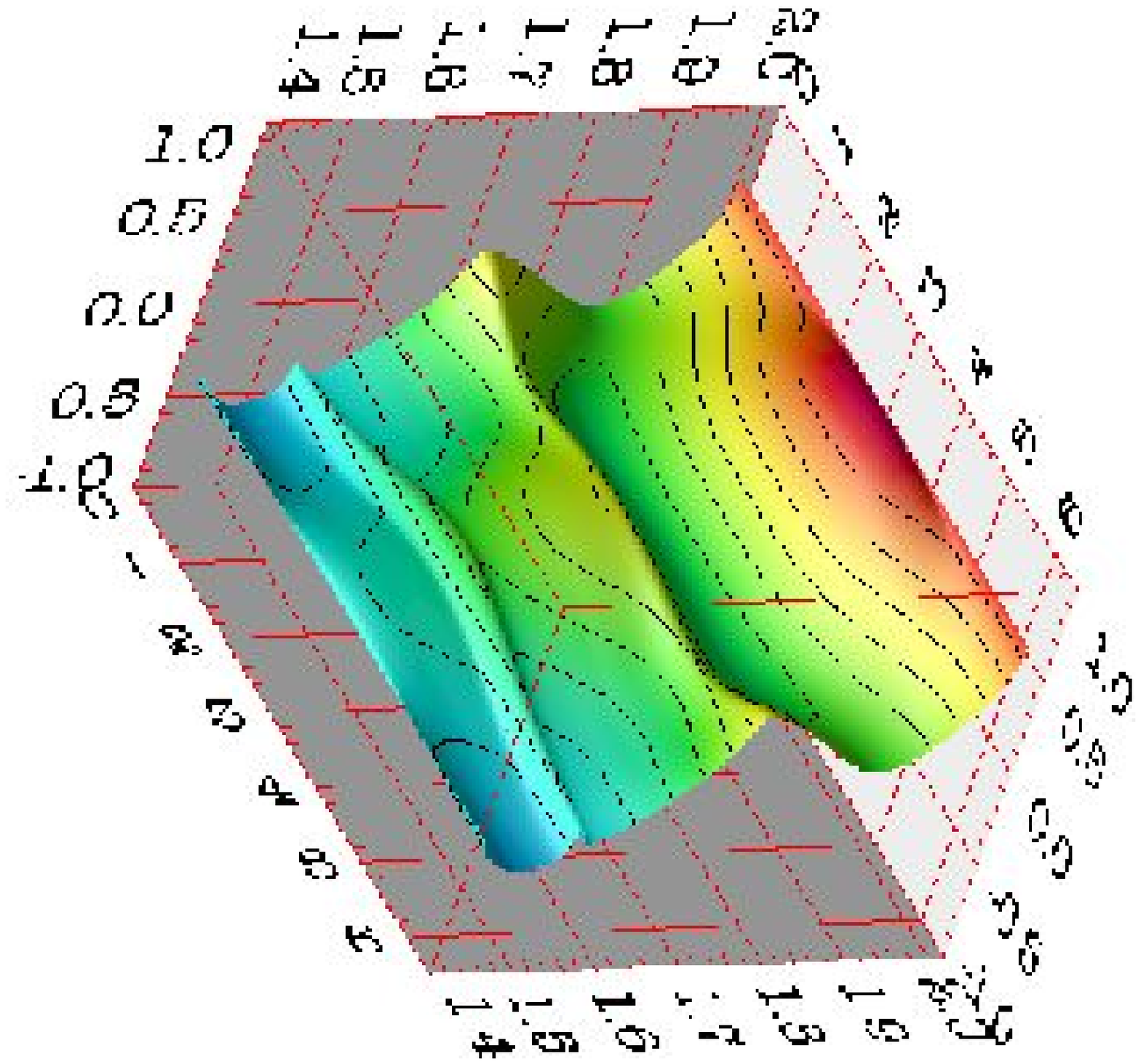}}
\end{figure}

\begin{figure}
\caption{The observable $P^\odot_z$, showing its sensitivity to the parity of the exotic resonance 
$\Theta^+$. (a) results when a $\Theta^+$ having $J^P=1/2^+$ is included. (b)
results when this state has $J^P=1/2^-$. Both surfaces are shown as 
functions of $m_{NK}$ and $\Phi^*$.\label{thetaf}}
\vskip 0.2in
\centerline{(a)\includegraphics[height=4.0in,angle=0]{pol_005_spp1_19.ps}}
\vskip0.2in
\centerline{(b)\includegraphics[height=4.0in,angle=0]{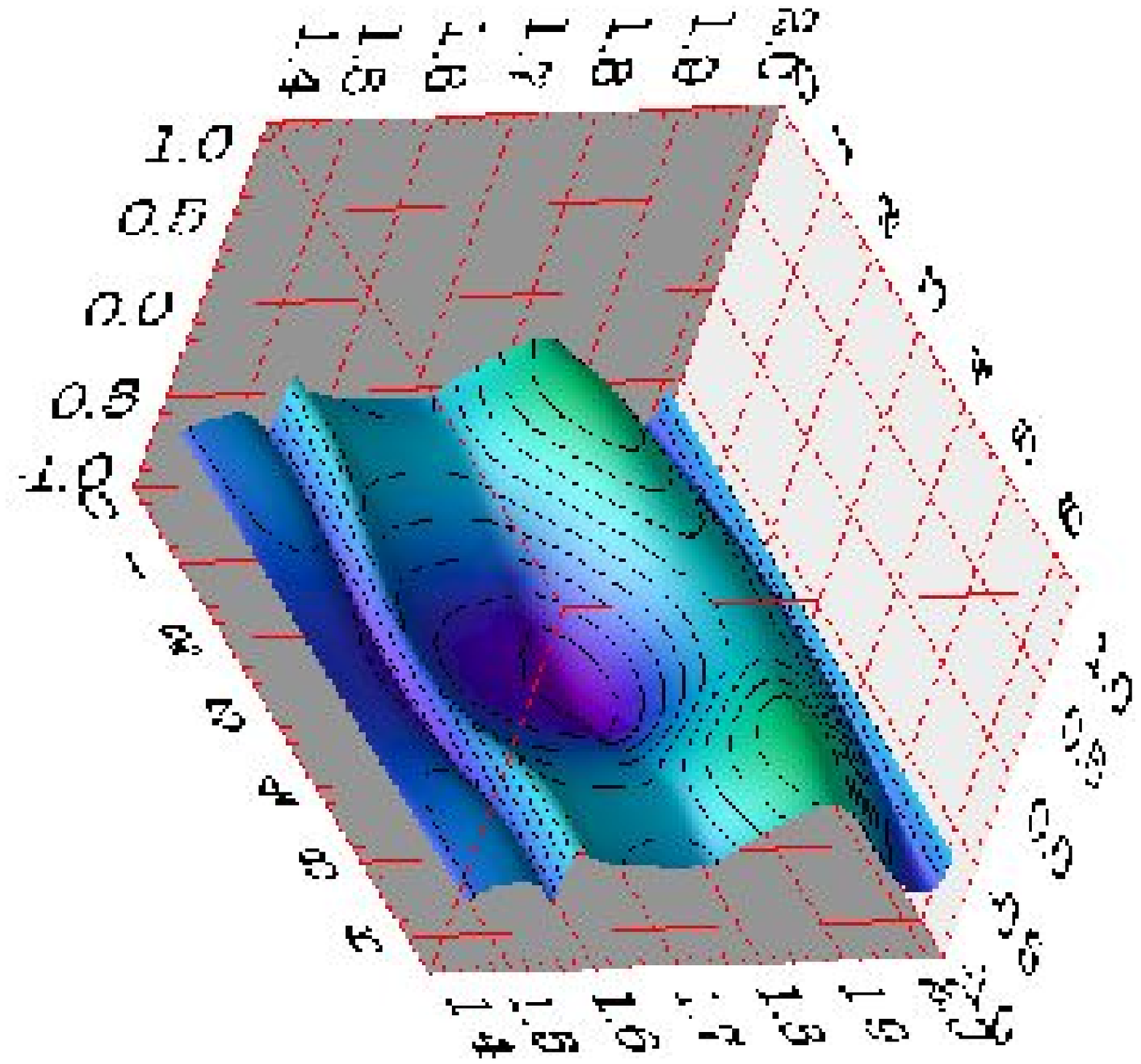}}
\end{figure}

Figs. \ref{thetad} to \ref{thetaf} ($I^\odot$, $P_x^\odot$ and $P_z^\odot$,
respectively) show the surfaces that result when the $\Theta^+$ has $J^P=1/2^+$
(a), compared with when it has $J^P=1/2^-$ (b), for the process $\gamma p\to
nK^+\overline{K}^0$. In fig. \ref{thetad}, there is a marked difference in the
surfaces that result, but for this observable, a pentaquark of negative parity
would be very difficult to isolate, as the `signal' is marginally different from
the background.

Fig. \ref{thetae} and \ref{thetaf} show more `detectable' signals for the
negative parity $\Theta^+$, and in each case the signal is noticably
different from that given by a $\Theta^+$ with positive parity. We emphasize
that while `visual' examination of any data obtained may provide hints at
underlying dynamics (such as the existence or not of the $\Theta^+$), detailed
comparison with the predictions of a model of some sort will be needed to
provide more concrete, quantitative and trustworthy interpretations.

In the above, we have chosen three observables to illustrate how it might be
possible to use these polarization observables in pentaquark searches. The
three observables we chose all required circularly polarized photons. It must
be pointed out here that all 63 observables show some kind of effect due to the
pentaquark, and some of the effects are quite striking. 

\section{Conclusion and Outlook}

The preceding picture show should have conveyed a number of points about the
polarization observables developed in \cite{polarization}. The first point is
that these observables may be displayed in a number of ways. The second, and
perhaps most obvious point to note is that however they are displayed, these
observables exhibit an enormously rich structure, reflecting the degree of
complexity in the underlying dynamics. This sensitivity to the various
contributions leading to the final state being studied, especially to `small'
contributions, provides an indispenable tool that will need to be fully exploited in
our attempts to understand processes like the ones discussed herein. Such
processes are expected to be among the primary sources of information required
in the on-going attempts to understand the dynamics of soft QCD.

As we have mentioned before, a number of these observables should be accessible
in the near future at existing facilities, in a number of different processes.
The obvious applications are to the process discussed herein, $\gamma N \to
NK\overline{K}$, and to $\gamma N\to N\pi\pi$. However, final states like
$N\eta\pi$, $N\eta\eta$, $YK\pi$ (where $Y$ is a $\Lambda$ or $\Sigma$), 
$YK\eta$, and even $KK\Xi$, will require the same kinds of measurements in
order to disentangle the various contributions leading to them. In the
processes that produce hyperons in the final states, their various
self-analysing decays provide access to recoil polarization measurements, thus
opening up more possibilities. Many of these opportunities will have to be
seized for continued progress to be made in our understanding of baryon
spectroscopy.

\vspace*{0.5in}

\begin{thebibliography}{99}

\bibitem{polarization} W. Roberts and T. Oed, nucl-th/0410012, submitted to Physical
Review C.

\bibitem{zplus} W. Roberts, nucl-th/0408034, to appear in Physical Review C.

\bibitem{strauch} S.~Strauch [CLAS Collaboration], arXiv:nucl-ex/0407008;
S.~Strauch [CLAS Collaboration], in proceedings of 2nd Conference on  Nuclear
and Particle Physics with CEBAF at JLab (NAPP 2003), Dubrovnik,  Croatia, 26-31
May 2003; manuscript in preparation.

\bibitem{penta} T. Nakano {\it et al.} [LEPS Collaboration],
        Phys. Rev. Lett. {\bf 91}, 012002 (2003);\\
V. Barmin {\it et al.} [DIANA Collaboration],
Phys.\ Atom.\ Nucl.\  {\bf 66}, 1715 (2003)
[Yad.\ Fiz.\  {\bf 66}, 1763 (2003)];\\
S. Stepanyan {\it et al.} [CLAS Collaboration],
Phys.\ Rev.\ Lett.\  {\bf 91}, 252001 (2003);\\
J. Barth {\it et al.} [SAPHIR Collaboration],
Phys.\ Lett.\ B {\bf 572}, 127 (2003);\\
V. Kubarovsky {\it et al.} [CLAS Collaboration],
Phys.\ Rev.\ Lett.\  {\bf 92}, 032001 (2004)
[Erratum-ibid.\  {\bf 92}, 049902 (2004)];\\
A. E. Asratyan, A. G. Dolgolenko, and M. A. Kubantsev,
Phys.\ Atom.\ Nucl.\  {\bf 67}, 682 (2004) [Yad.\ Fiz.\  {\bf 67}, 704 (2004)];\\
A. Airapetian {\it et al.} [HERMES Collaboration],
Phys.\ Lett.\ B {\bf 585}, 213 (2004);\\
A. Aleev {\it et al.} [SVD Collaboration],
	hep-ex/0401024;\\
M. Abdel-Bary {\it et al.} [COSY-TOF Collaboration],
	hep-ex/0403011.


\end{thebibliography}
\end{document}